\documentclass[journal]{IEEEtran}
\usepackage{amsmath,amsfonts}
\usepackage{algorithmic}
\usepackage{algorithm}
\usepackage{array}
\usepackage[caption=false,font=normalsize,labelfont=sf,textfont=sf]{subfig}
\usepackage{textcomp}
\usepackage{stfloats}
\usepackage{url}
\usepackage{verbatim}
\usepackage{graphicx}
\usepackage{cite}

\usepackage{siunitx}
\usepackage{physics}

\begin{document}

\title{Load Balancing in Strongly Inhomogeneous Simulations -- a Vlasiator Case Study}




\author{Leo Kotipalo,
\thanks{
This work has been submitted to the IEEE for possible publication. Copyright may be transferred without notice, after which this version may no longer be accessible.
}%
\thanks{%
    This article was based on code development part of the EuroHPC Joint Undertaking Plasma-PEPSC project, grant 4100455.
    This article was supported by the Research Council of Finland grants 
    328893, 
    336805, 
    339756, 
    345701, 
    347795, 
    352846, 
    353197, 
    359806  
    and 361901 
    and European Research Council grant 101124500-WAVESTORMS.
    The Finnish Centre of Excellence in Research of Sustainable Space, funded through the Research Council of Finland grants 312351 and 336805, has significantly supported Vlasiator development, as has the European Research Council Consolidator Grant 682068-PRESTISSIMO.
    \emph{(Corresponding author: Leo Kotipalo.)}
}%
\thanks{L. Kotipalo (leo.kotipalo@helsinki.fi), M. Battarbee, and V. Tarvus are with the Faculty of Science, University of Helsinki, Helsinki, Finland}
Markus Battarbee,
Yann Pfau-Kempf,
\thanks{Y. Pfau-Kempf is with CSC -- IT Center for Science Ltd., Espoo, Finland}%
Vertti Tarvus,
Minna Palmroth
\thanks{M. Palmroth is with the Faculty of Science, University of Helsinki, Helsinki, Finland, and also with the Finnish Meteorological Institute, Helsinki, Finland}%
\thanks{This article was produced in collaboration with the University of Helsinki space physics research group.}%
\thanks{Data is available on-line at https://doi.org/10.23729/fd-9e84a409-0da9-3223-ad8b-011fd6112ce3}%
}



\maketitle

\begin{abstract}
Parallelization is a necessity for large-scale simulations due to the amount of data processed.
In this article we investigate different load balancing methods using Vlasiator,
a global magnetospheric simulation as our case study.

The theoretical basis for load balancing is the (hyper)graph partitioning problem,
modeling simulation units as vertices and their data dependencies as edges.
As it is an NP-hard problem, heuristics are necessary for dynamic runtime balancing.

We consider first hypergraph partitioning via an algorithm called parallel hypergraph partitioner (PHG);
this is done by partitioning a simplified grid and then attempting to optimize the solution on the finer grid.
The second and third are the geometric methods of recursive coordinate bisection (RCB) and recursive inertial bisection (RIB).

Finally we consider the method of Hilbert space filling curves (HSFC).
The algorithm projects simulation cells along a Hilbert curve and makes cuts along the curve.
This works well due to the excellent locality of Hilbert curves,
and can be optimized further by choice of curve.
We introduce and investigate six three-dimensional Hilbert curves in total.

Our findings on runs of two different scales indicate the HSFC method provides optimal load balance,
followed by RIB and PHG methods and finally by RCB.
Of the Hilbert curves evaluated,
the Beta curve outperformed the most commonly used curve by a few percent.
\end{abstract}

\begin{IEEEkeywords}
Load balancing, Hilbert curves, Plasma simulation
\end{IEEEkeywords}

\section{Introduction and theory} \label{sec:intro}
\IEEEPARstart{D}{evelopment} in computing performance is increasingly focused on parallelization,
and balancing the load between processes is imperative for any high-performance application.
A typical domain-decomposed simulation has each process solve a part of the problem in its own domain,
followed by communication of solved values at domain interfaces between the processes.
If a process has significantly less load than its neighboring processes, it will have to wait while the neighbors are still calculating, wasting computational resources.

\begin{figure*}[!t]
    \centering
    \includegraphics[width=6in]{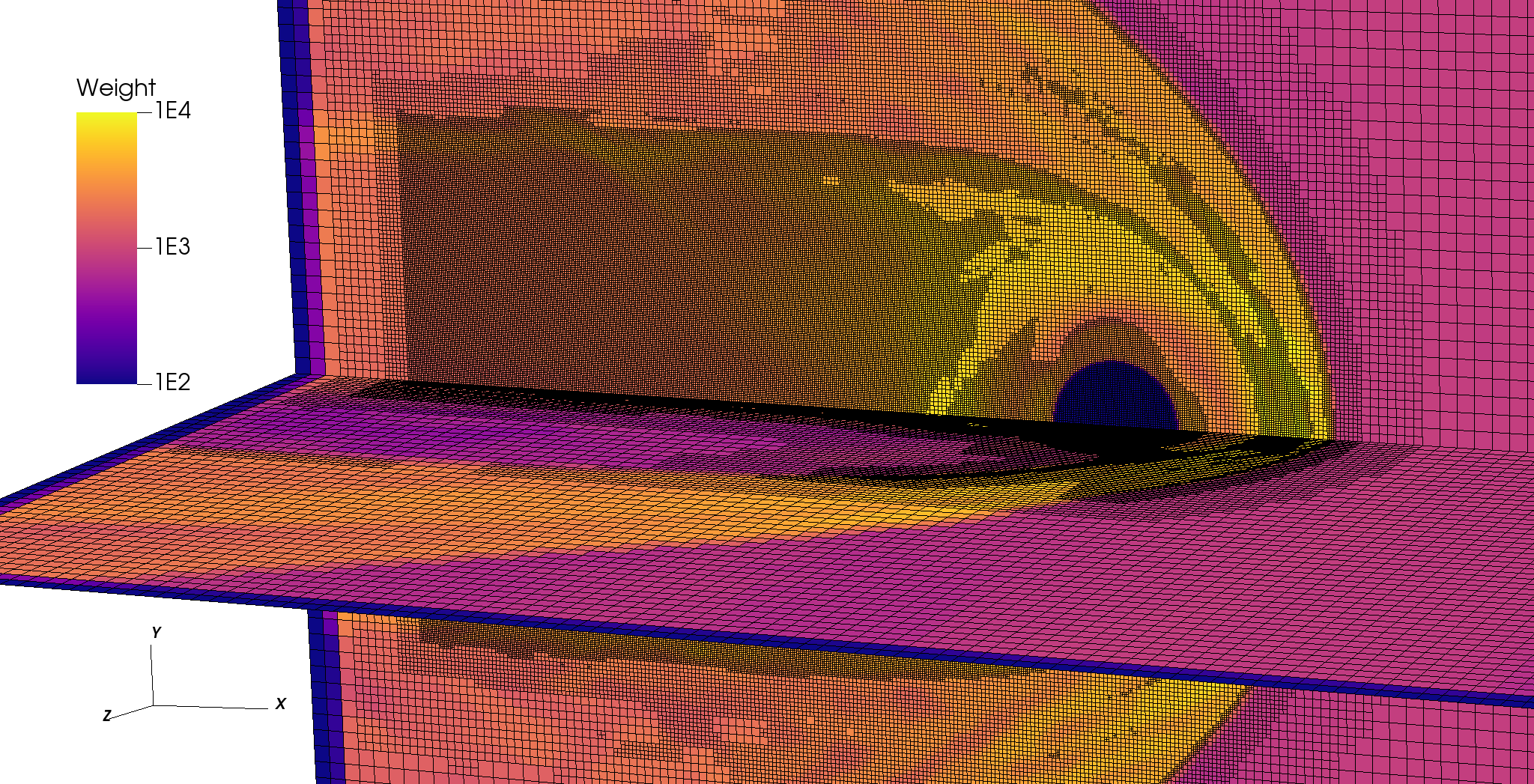}
    \caption{
        Load balancing weights (color on a logarithmic scale) in a three-dimensional Vlasiator run, two slices with grid (black cubic mesh) overlaid.
        As can be seen, memory and computational load per cell can vary by two orders of magnitude in the simulation region;
        additionally spatial resolution can be up to eight times finer (three levels of cell-based octree refinement) in areas of interest.
    }
    \label{fig:lbweight-twoslice}
\end{figure*}

Our research question is the evaluation of load balancing algorithms in strongly inhomogeneous simulations, using the global hybrid-Vlasov magnetospheric model Vlasiator \cite{vlasiatorpaper,vlasiator531} as the test case. 
Vlasiator is a particularly challenging simulation to balance.
As the simulation domain is physically inhomogeneous,
cubic simulation cells are implemented with both variable spatial resolution and varying amounts of simulated velocity space cells.
An example run is in Figure \ref{fig:lbweight-twoslice},
demonstrating the scale of difference in computational load between different regions.
As load balancing weight can vary by two orders of magnitude in the simulation,
so too can the amount of cells per process.
With the addition of adaptive mesh refinement, the largest subdomain can be thousands of times larger in physical volume than the smallest.
Load balancing is provided by the library Zoltan \cite{zoltan}.

The structure of the article is as follows. 
Section \ref{sec:intro} describes the theoretical background of load balancing.
Section \ref{sec:algs} goes through algorithms used to approximate the problem, ranging from graph-based to geometric methods with specific attention given to the optimization of Hilbert curves.
Section \ref{sec:software} describes the specifics of Vlasiator and the Zoltan load balancing library.
Section \ref{sec:methods} describes the test set-up used to evaluate load balancing performance.
Section \ref{sec:results} gives the results of the tests done, comparing both the different algorithms and different curves used for Hilbert curve partitioning.
\IEEEpubidadjcol 

\subsection{Theory} \label{sec:theory}
A simple way of modeling load balancing is apportioning cells to processes according to just their computational weight, i.e. the time taken to solve the cell.
This is called \emph{multiway number partitioning} and is an NP-hard problem \cite{Karp1972, np-completeness}.
As simulation time ideally scales linearly with cells simulated,
it is thus unfeasible to determine the exact solution with dynamic runtime load balancing.
However, good approximations exist.
To evaluate the quality of a solution, 
we can take the ratio of the largest weight partition in the solution to the largest weight in an ideal solution;
this is called the \emph{approximation ratio} $\alpha_\mathrm{w}$.
Greedy partitioning, where we iterate over cells and place each in the process with the least weight currently,
is a linear time solution with a worst case approximation ratio of $2$ for an arbitrary amount of partitions.
If we sort the cells first by weight in $\order{n_\mathrm{v} \log n_\mathrm{v}}$ time for $n_\mathrm{v}$ vertices,
the ratio is improved to $4/3$ \cite{lptscheduling}.

\begin{figure}[!t]
    \centering
    \includegraphics[width=3in]{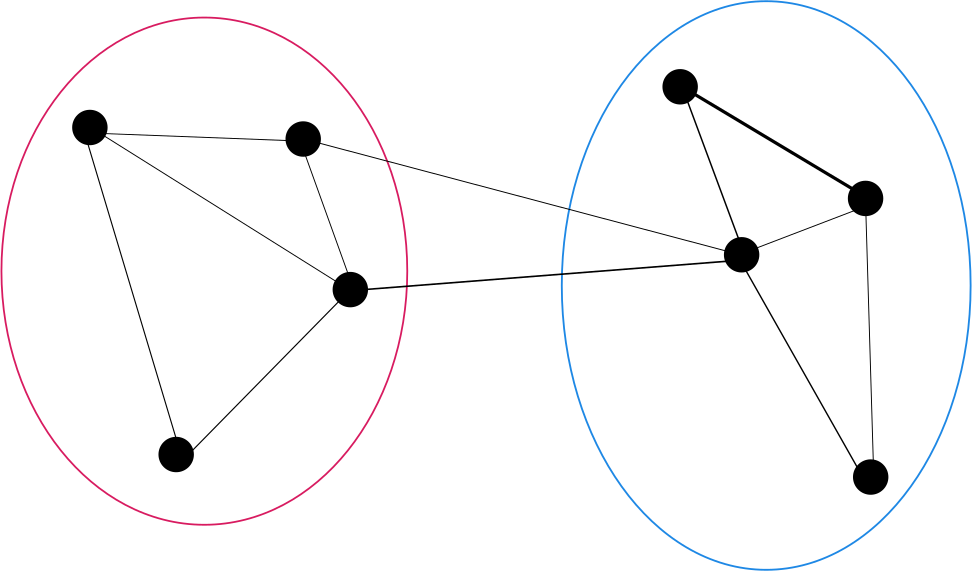}
    \caption{
        Illustration of graph partitioning.
        The objective is to partition the vertices of a weighted graph into disjoint parts such that the vertex weights are balanced,
        and the weights of cut edges, edges whose vertices are in different partitions,
        are minimized. 
        Here we assume uniform weights for the vertices and the edges, respectively.
    }
    \label{fig:graph-partition}
\end{figure}

Unfortunately this simple model is only applicable to the most trivially parallelizable applications where no communication is required.
In a typical stencil-based simulation, each cell requires data from nearby cells for propagation;
on process boundaries, this is handled by \emph{ghost cells}, local copies of relevant data from cells in other processes.
Partitioning only according to weight doesn't account for communication, and disconnected, distorted or concave domains result in excessive communication and ghost domain memory footprint.
Instead we may consider the cells as vertices in a graph, and the communication dependencies between them as the edges of the graph.
Vertices are weighted according to the computation weight as in number partitioning, and edges according to the total cost of data communication in situations where the vertices it connects are on different processes.
We now want to divide the graph into components such that the vertex weights are balanced, as in number partitioning.
Additionally the cut size, the total weight of edges whose vertices are in different partitions, should be minimized.
Similarly to number partitioning, we may evaluate an approximate solution using a vertex weight approximation ratio $\alpha_\mathrm{v}$ and a edge cut approximation ratio $\alpha_\mathrm{e}$.
$\alpha_\mathrm{v}$ is defined as $\alpha_\mathrm{w}$ in number partitioning, 
and $\alpha_\mathrm{e}$ is the ratio of the cut size of the approximate solution to the cut size in the ideal solution.
This is called \emph{graph partitioning}, which is also NP-hard \cite{graph-partition}.
An illustration is in Figure \ref{fig:graph-partition}.

Graph partitioning can also be extended to \emph{hypergraphs}.
These are similar to graphs but instead of edges connecting two vertices,
hypergraphs have \emph{hyperedges} connecting an arbitrary number of vertices \cite{Berge}.
Hypergraphs can model the communication volume of certain applications better than traditional graphs \cite{hypergraph-partitioning}.
Note that graphs are effectively degenerate hypergraphs in the case where each hyperedge's degree, the amount of vertices connected, is two.

For practical simulation purposes a \emph{near-balanced} solution in vertex weights is necessary,
for which the weight approximation ratio is at most $\alpha_\mathrm{v} = 1 + \epsilon$ for some arbitrary $\epsilon > 0$.
Combining this requirement with minimizing the edge cut approximation ratio $\alpha_\mathrm{e}$ is difficult.
In the specific case of two-dimensional graphs modeling a grid of cells with edges connecting face neighbors,
there is no near-balanced polynomial time algorithm for which $\alpha_e = n_\mathrm{v}^c / \epsilon^d$ for vertex count $n_\mathrm{v}$, vertex weight imbalance $\epsilon$ and arbitrary parameters $c, d < 1/2$ unless $\mathrm{P}=\mathrm{NP}$ \cite{fastbalancedpartitioninghard}.
As such, heuristic methods described in Section \ref{sec:algs} have no guarantees on the cut size.

\section{Load balancing algorithms} \label{sec:algs}
Several heuristic methods exist for load balancing.
We may loosely divide them into algorithms for near-balanced graph partitioning such as PARMETIS \cite{parmetis} and PT-Scotch \cite{pt-scotch},
and geometric methods taking advantage of physical simulations typically having a strong connection between the physical location of simulation elements and their communication dependencies.
We've constrained our analysis to algorithms supported natively by Zoltan \cite{ZoltanDevelopersGuideV3}.
Algorithms excluded are the graph partitioners PARMETIS and PT-Scotch supported by but not included in Zoltan,
refinement tree partitioning \cite{REFTREE} due to Zoltan's implementation not supporting 64-bit IDs used by Vlasiator,
and the simple partitioners Block, Cyclic and Random due to their extremely poor performance.

Another notable exclusion is \emph{hierarchical} partitioning,
which allows two different partitioning methods to be combined such that the partition produced by one partitioner can be subdivided by another.
In theory this could allow optimizing for communication on inter-node partitioning and computational load for intra-node partitioning.
We decided against analyzing it in this article for two reasons;
the first being the relatively small difference between intra-node and interconnect bandwidth on the LUMI hardware used for testing of \SI{256}{GB\per\s} and \SI{200}{GB\per\s} respectively \cite{lumihw},
and the second being that testing every combination of hierarchical partitions would increase the amount of trials substantially.

\subsection{Hypergraph partitioning}
\begin{figure*}[!t]
    \centering
    \subfloat[]{
        \includegraphics[width=3in]{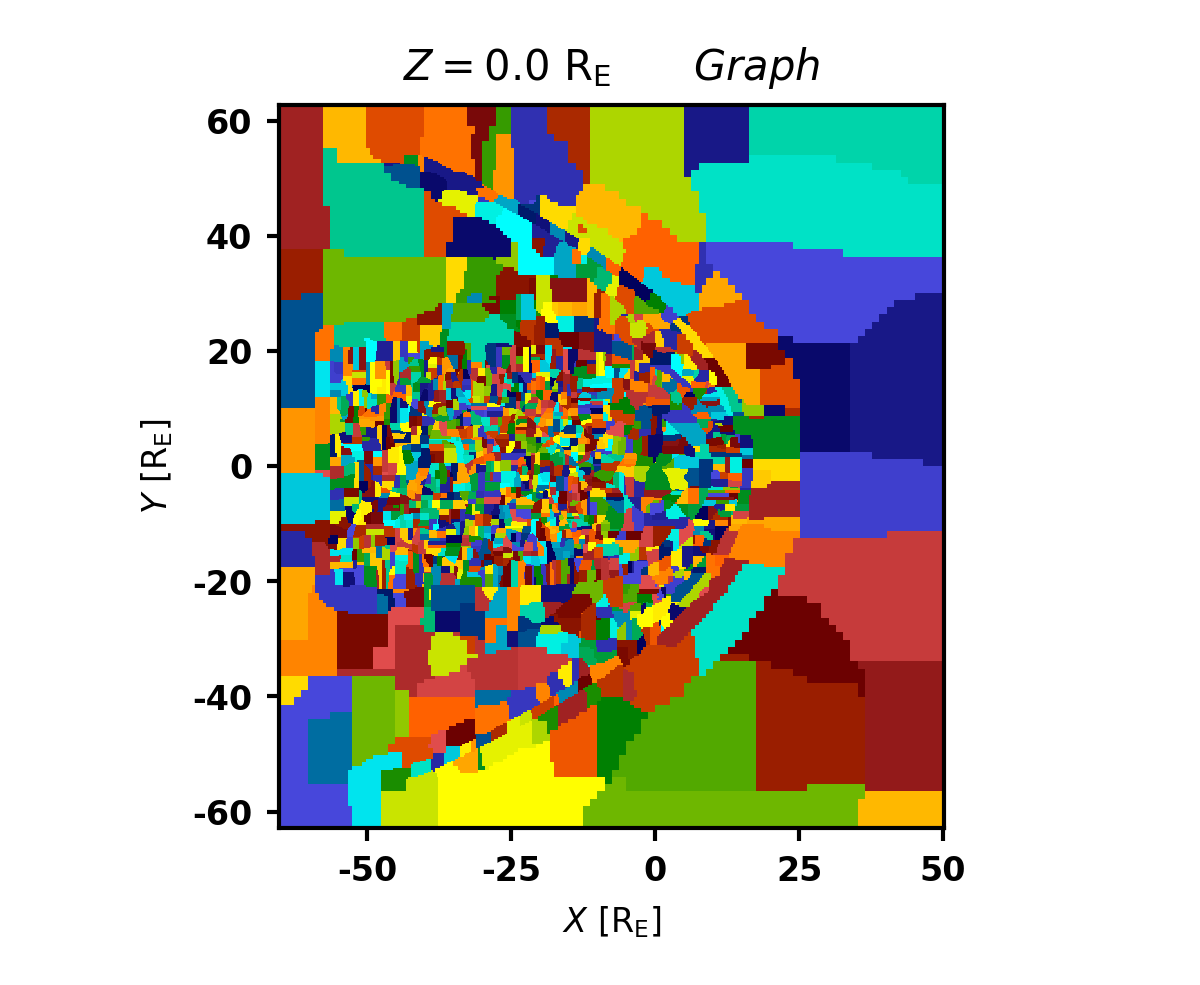}
    } \hfil
    \subfloat[]{
        \includegraphics[width=3in]{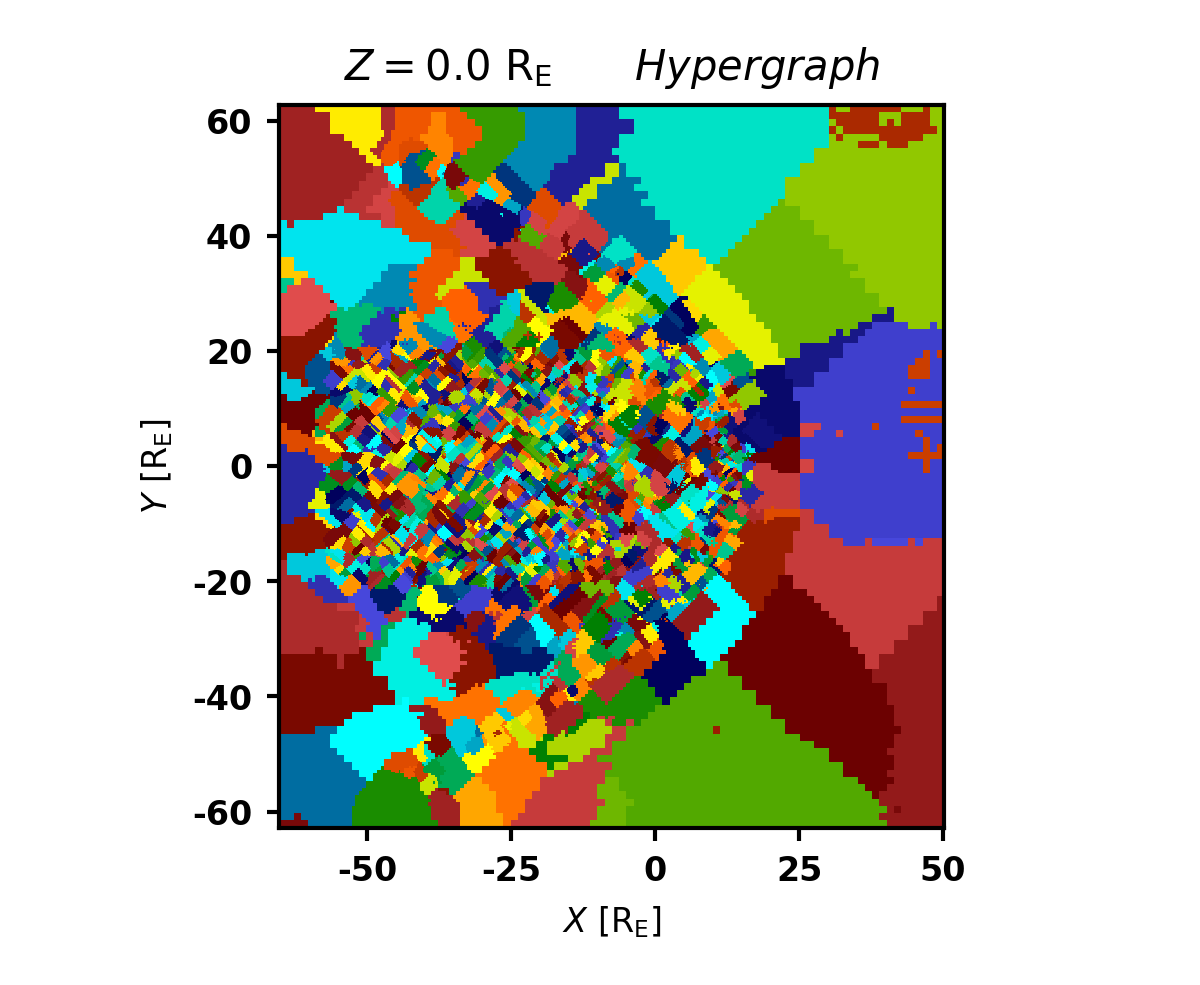}
    }
    \caption{
        Examples of PHG graph (a) and hypergraph (b) partitioning, one slice of a three-dimensional simulation with \num{8000} processes.
        Domains are colored by process number modulo 64 to increase contrast between neighboring domains.
        Computationally heavy regions can be discerned by the smaller domain sizes, an example being $x < \SI{25}{R_E}$, $\abs{y} < \SI{20}{R_E}$.
        The seemingly discontinuous regions seen around $x > \SI{25}{R_E}$ in Hypergraph are due to this being a two-dimensional slice; these are likely part of a domain on another plane.
    }
    \label{fig:phg}
\end{figure*}

As established in Section \ref{sec:theory}, optimal solutions for graph partitioning and by extension hypergraph partitioning are impossible in polynomial time.
Here we will consider the \emph{Parallel Hypergraph partitioner} (PHG) \cite{phg}.
PHG partitions hypergraphs recursively, splitting the subdomains in half until the domain is partitioned to a desired amount of parts.
The workaround used by PHG is to decrease the size of the hypergraph by merging vertices based on the amount of hyperedges they share; 
this is called coarsening.
Once the hypergraph is small enough, the graph is partitioned in half using global optimization.
This partitioning is then projected to the original hypergraph, and refined via local optimization, moving vertices between subdomains such that every change improves the cut weight \cite{phg}.
As graphs are degenerate hypergraphs, PHG can be used for graph partitioning as well.
Examples of this algorithm for both graph and hypergraph partitioning, as applied to the Vlasiator simulation, are provided in Figure \ref{fig:phg}.

\subsection{Recursive bisection} \label{sec:rec}
Instead of explicitly considering the (hyper)graph of cells and connections,
we may take advantage of geometric methods for partitioning into well-balanced domains with minimal communication.
As the computational units in physical simulations are almost always cells or particles with positions,
we can use this information to our benefit in load balancing.
Since physical effects have a limited propagation speed,
simulation cells far away from each other don't interact directly during a single timestep.
This means a cell having nearby cells in the same partition is favorable,
while spatially spread out partitions perform comparatively worse.
More generally we can expect an ideal partition to be composed of shapes that have a large ratio of volume to surface area,
which roughly correspond to computational load and communication load respectively.

We consider here two recursive methods, \emph{recursive coordinate bisection} (RCB), also called orthogonal recursive bisection and \emph{recursive inertial bisection} (RIB), also called eigenvalue recursive bisection \cite{rib}.
Both methods recursively split the domain with different choices of cut plane, dividing computational weight in half between the sub-domains.

In RCB we first determine the ``computational center of mass'' of the domain via the computational weights and coordinates of cells.
Then, the domain is split into two through the center of mass along the longest coordinate axis.
This turns out to be a good heuristic for relatively uniform simulations;
rectilinear blocks are sufficient for a good ratio of volume to surface area,
which would typically correspond to reduced communication \cite{rcb}.
As the center of mass is calculated as a position in continuous space while simulation elements are often on a discrete grid,
we may choose to either make the cut strictly between grid cells enforcing rectilinearity and improving communication,
or to move individual cells between the two domains to improve the weight balance.

\begin{figure*}[!t]
    \centering
    \subfloat[]{
        \includegraphics[width=3in]{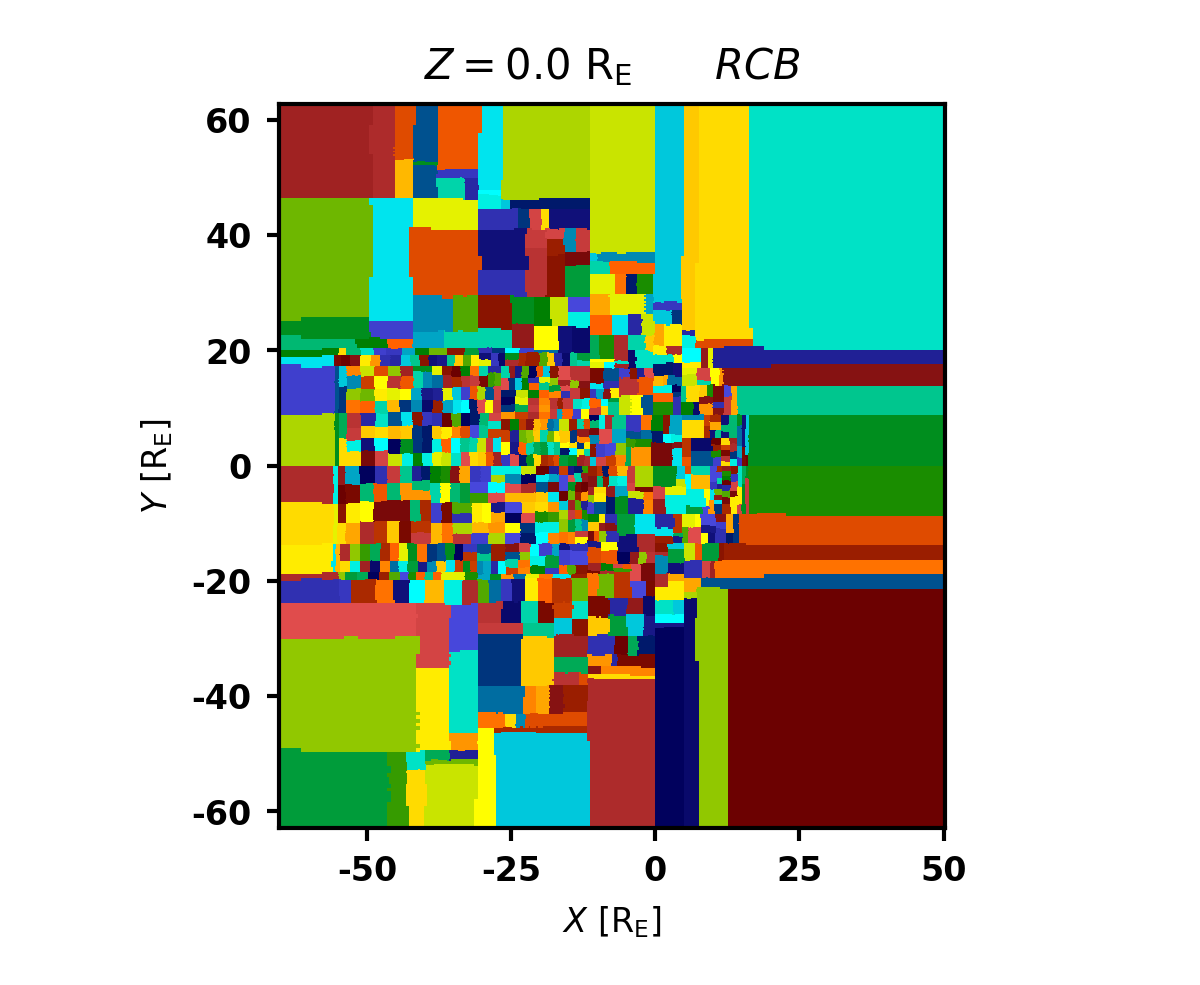}
    } \hfil
    \subfloat[]{
        \includegraphics[width=3in]{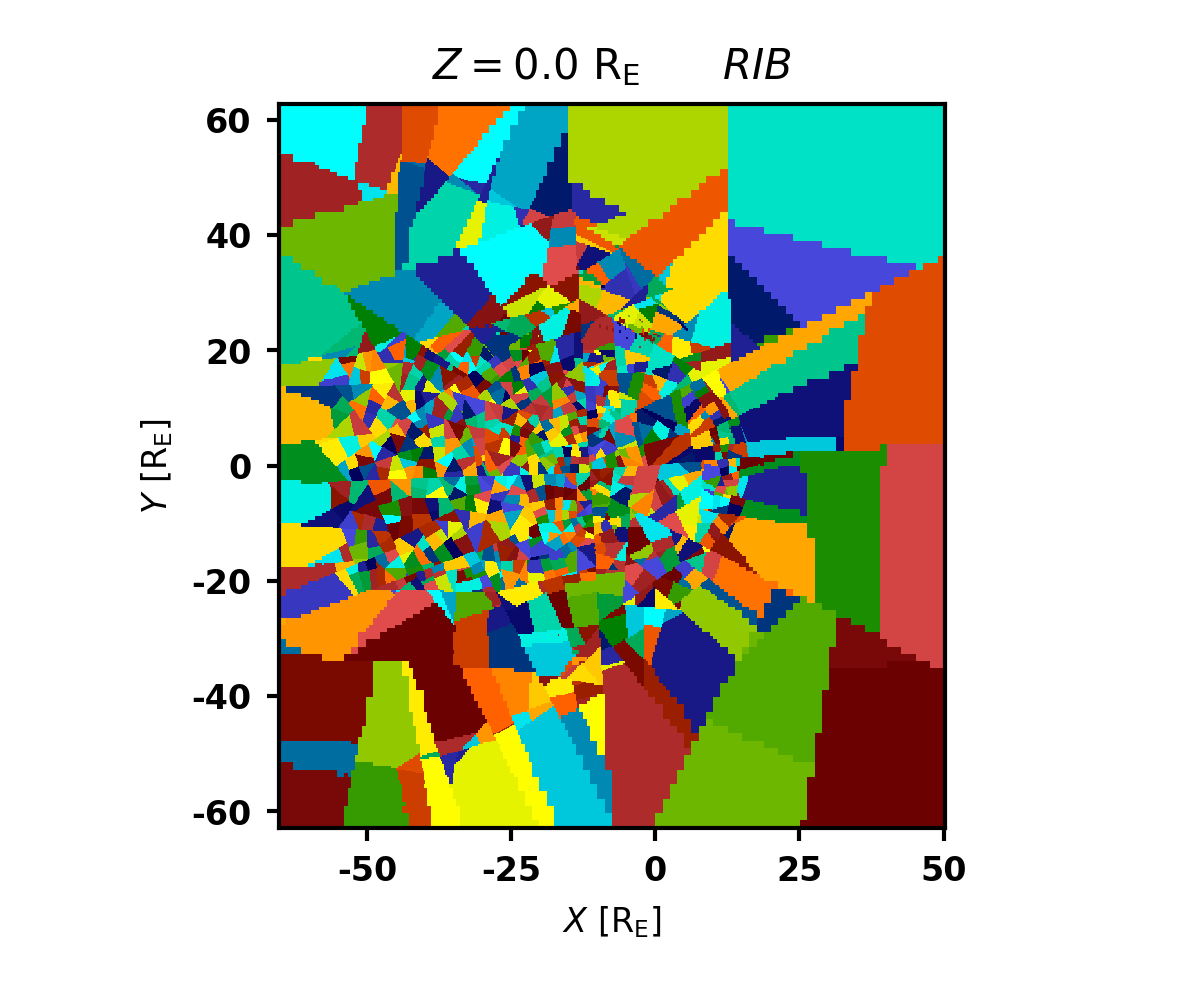}
    }
    \caption{
        Examples of recursive coordinate (a) and inertial bisection (b), one slice of a three-dimensional simulation with \num{8000} processes.
        Domains are colored by process number modulo 64.
        The difference between rectilinear blocks in RCB and oblique `shards' in RIB is clear.
        Note that the lines in RCB aren't entirely straight;
        here the weight balance is improved at the cost of rectilinearity.
    }
    \label{fig:bisections}
\end{figure*}

Similarly in RIB the cuts are done recursively through the center of mass.
However, the cut plane is determined by calculating the ``computational inertia tensor'' of the domain as if the computational weights were physical masses.
The cut is then done perpendicular to the axis of minimum inertia.
This effectively works to maximize the weighted distance of cells from the cut plane \cite{rib}.
Examples of both methods in Vlasiator are provided in Figure \ref{fig:bisections}, 
showing rectilinear blocks for RCB and oblique `shards' for RIB.
Note that some of the blocks in RCB are rather thin; this may cause issues in load balancing of communication along different spatial directions.

\subsection{Hilbert space filling curve} \label{sec:hsfc}
\begin{figure*}[!t]
    \centering

    \subfloat[]{
        \includegraphics[width=3in]{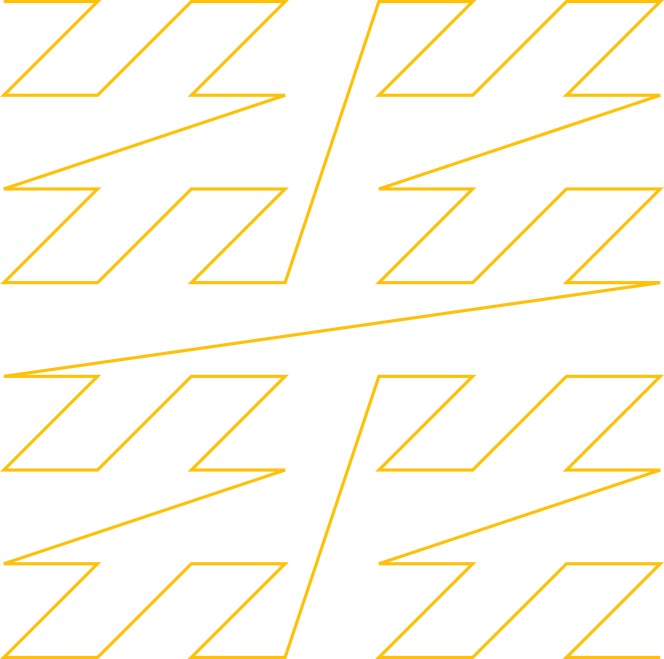}
    }
    \hfil
    \subfloat[]{
        \includegraphics[width=3in]{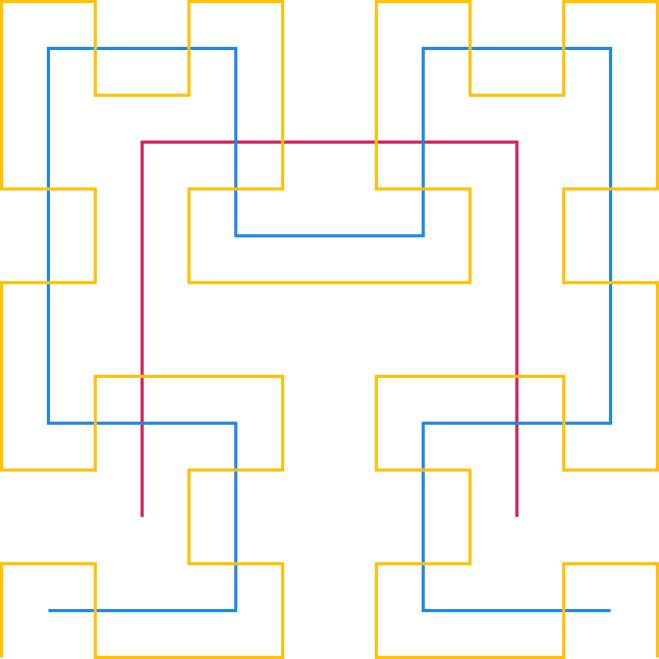}
    }
    \caption{
        The third order approximation of the two-dimensional Z-curve (a) and the first three orders of approximation of the two-dimensional Hilbert curve in fuchsia, cyan and yellow (b).
        Subsequent orders of approximation can be formed by replacing quadrants with a rotation of the first order curve.
    }
    \label{fig:space-filling-curves}
\end{figure*}

The final option considered in this article is the method of space filling curves.
Space-filling curves are one-dimensional fractal curves that range over a higher-dimensional shape such as a square or a cube,
forming a mapping from $n$-dimensional to one-dimensional space.
The simplest implementation is the Z-curve, formed by interleaving the bits of the coordinate values of a point \cite{morton};
in two dimensions, this forms a recursive Z-shape shown in Figure \ref{fig:space-filling-curves}(a).
In this load balancing method, the cells in the three-dimensional simulation domain are mapped to the curve, with partitioning done by making cuts along the curve.

The optimal choice is the \emph{Hilbert space-filling curve} (HSFC) \cite{hilbert} and its three-dimensional extensions,
as they have excellent properties for load balancing.
Hilbert curves are continuous and preserve locality -- points close to each other on the curve are close to each other in space \cite{haverkort},
meaning partitions along the curve end up connected.
It is worth noting that the opposite does not hold,
as the curve has `seams' at any point $x_i = 2^{-n}$ for any coordinate $x_i$ and $n$ up to the order of the curve, meaning there are points a finite distance away from each other on the curve with arbitrarily small distance in space.
Additionally as the curve is formed as an octant-to-octant traversal,
the partitions are roughly cuboidal, 
maximizing volume and minimizing surface and thus reducing communication load.
The first three orders of the two-dimensional Hilbert curve are shown in Figure \ref{fig:space-filling-curves}(b).
Subsequent orders are formed by replacing octants of the curve by the first order curve rotated and mirrored appropriately.


Spatial coordinates can be efficiently converted to Hilbert coordinates and vice-versa using bit-wise operations.
Essentially, the coordinates of cells are processed by taking the leading bit of each coordinate to determine the octant of space we are working in. 
This then corresponds to the leading three bits (eighth) of the Hilbert coordinate
and a \emph{state transformation} related to rotation and mirroring of the fractal curve.
This is then repeated with the next bit to determine the sub-octant and the next three bits of the Hilbert coordinate, and so on until the desired accuracy is reached \cite{BUTZpaper}.

\begin{figure}[!t]
    \centering
    \includegraphics[width=3in]{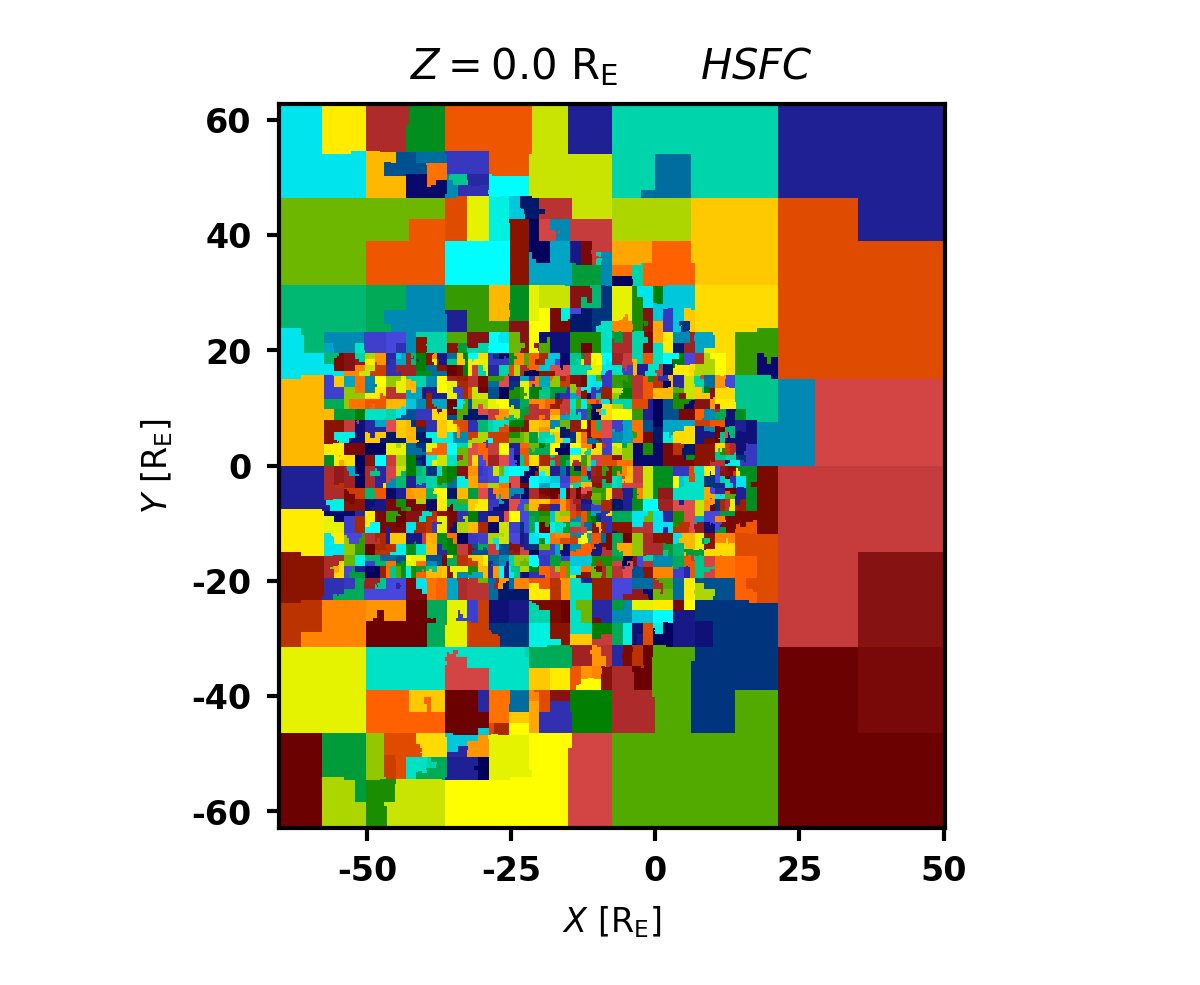}
    \caption{
        Example of Hilbert space filling curve partitioning using Zoltan's default HSFC curve (Octree), one slice of a three-dimensional simulation with \num{8000} processes.
        Domains are colored by process number modulo 64.
        Due to the Hilbert curve visiting each point in an octant before moving on to the next,
        the domain shapes are formed from consecutive octants with sub-octants from the previous and following octants.
        A marked difference from RCB is the tendency of domains to have approximately equal size in all dimensions.
    }
    \label{fig:HSFC-run}
\end{figure}

After determining the Hilbert coordinates of the cells they can be binned along the curve using a greedy algorithm, placing a cut when the cumulative sum of weights is greater than the target weight.
This partitioning is then refined by moving cells between bins \cite{ZoltanDevelopersGuideV3}.
An example from Vlasiator is provided in Figure \ref{fig:HSFC-run}.

\subsection{Hilbert curve optimization and pessimization} \label{sec:curve-optimization}
All space-filling curves are not created equal.
In two dimensions, \emph{the} Hilbert curve is uniquely defined as a self-similar continuous quadrant-to-quadrant traversal where consecutive quarters share a side.
This is the optimal choice for locality, compared to e.g. the simpler Z-curve.
In three dimensions, the problem is more complicated and there isn't even a straightforward definition of a Hilbert curve.
Extending the two-dimensional definition to octant-to-octant traversals, \num{10694807} curves exist, with varying properties \cite{haverkort}.
The primary properties of interest for load balancing are the locality metrics, specifically $L_p$-dilation $\mathrm{WL}_p$:
\begin{equation}
    \mathrm{WL}_p = \mathrm{max} \qty{\frac{D_p \qty(\tau(a), \tau(b))}{b - a}: 1 \geq b > a \geq 0}
\end{equation}
where $D_p (x, y) = \qty(\sum_i \left|x_i - y_i \right|^p )^{1/p}$ is the $L_p$-distance, $a$ and $b$ are two Hilbert coordinates, and $\tau(a)$ and $\tau(b)$ are the corresponding three-dimensional coordinates.
Effectively this is a ratio of distance in 3D-space divided by distance in the Hilbert coordinate, so low dilation means points that are far away will be far away on the curve.
Points far away would be less likely to end up on the same partition, 
which should reduce the amount of ghost cells in each domain and thus communication required.
The relevant distance metric depends on the communication patterns of the program.
If we consider cell-based simulations communicating data along coordinate directions,
we'd expect for $L_1$-dilation to be the most important for performance,
and our hypothesis is that lowering $L_1$-dilation will thus improve performance for Vlasiator.

\begin{figure*}[!t]
    \centering
    \subfloat[]{
        \includegraphics[width=2in]{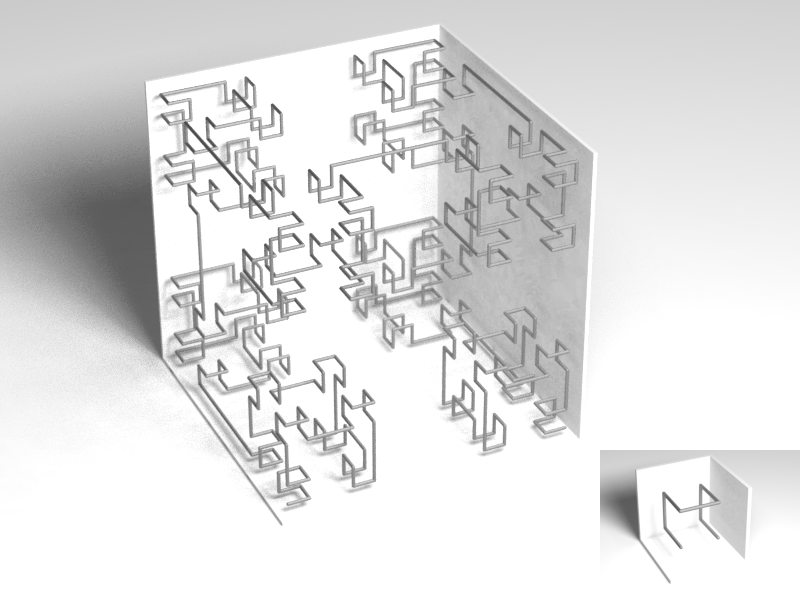}
    } \hfil
    \subfloat[]{
        \includegraphics[width=2in]{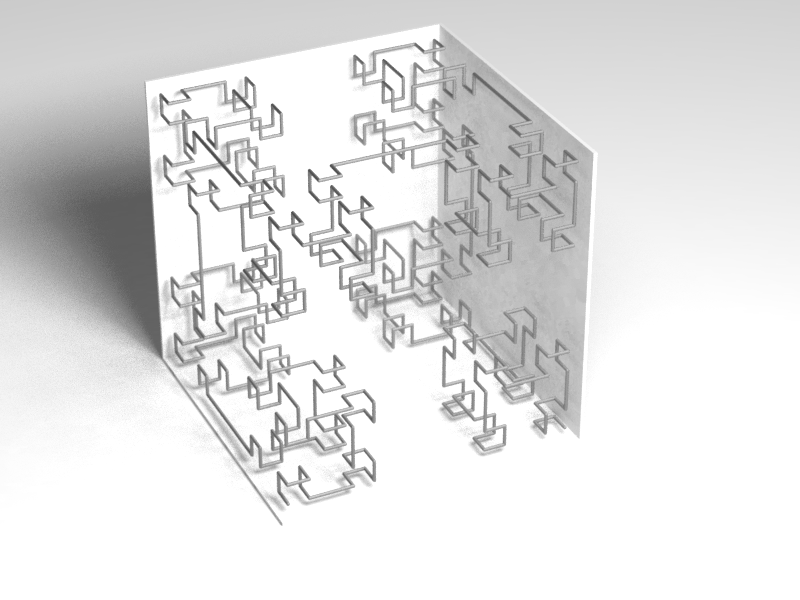}
    } \hfil
    \subfloat[]{
        \includegraphics[width=2in]{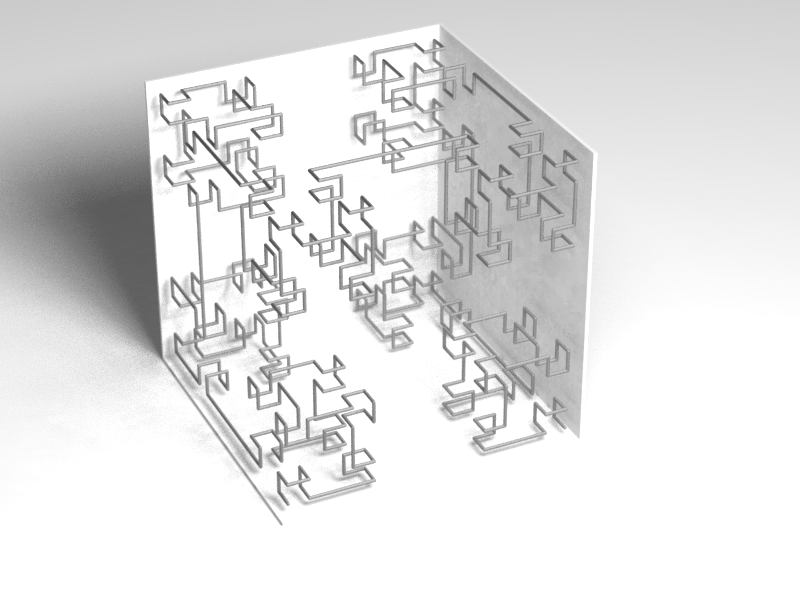}
    } \\
    \subfloat[]{
        \includegraphics[width=2in]{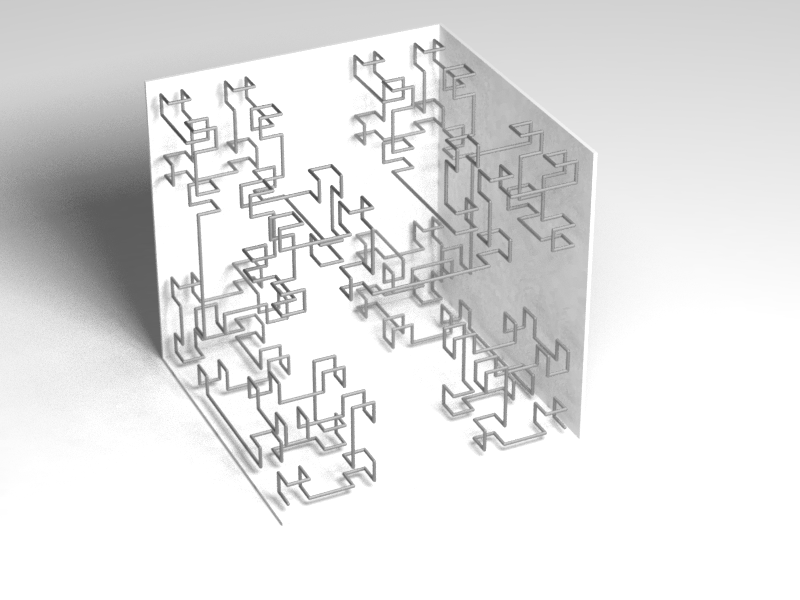}
    } \hfil
    \subfloat[]{
        \includegraphics[width=2in]{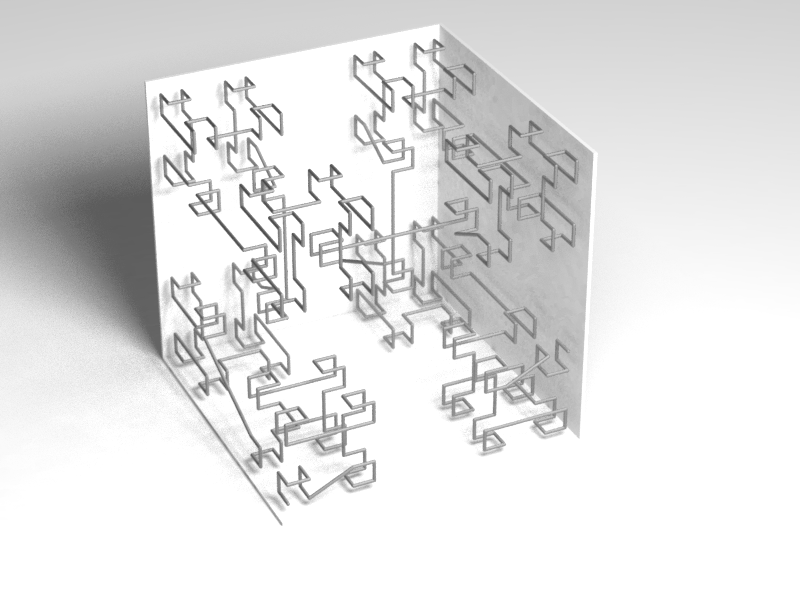}
    } \hfil
    \subfloat[]{
        \includegraphics[width=2in]{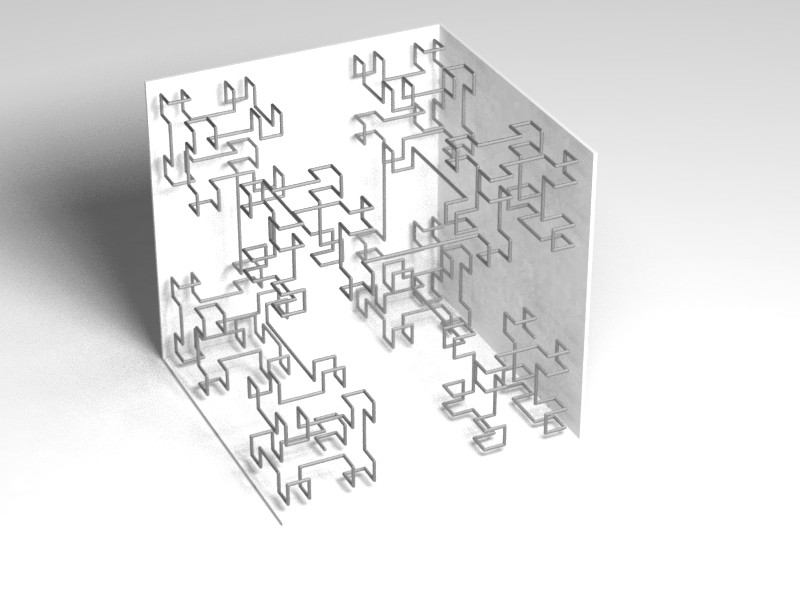}
    }
    \caption{
        The six three-dimensional Hilbert curves evaluated: Butz (a), Alfa (b), Harmonious (c), Sasburg (d), Base Camp (e) and Beta (f). 
        The grid here is a $8 \times 8 \times 8$ grid with distances between larger octants exaggerated for visual clarity.
        The base pattern Ca00 shared by these curves is the saddle shape shown in the inset of panel (a), which is repeated on all structural levels.
        Figures created with Hilbex \cite{haverkort} and POV-Ray \cite{povray}.
    }
    \label{fig:3d-hilberts}
\end{figure*}

The most commonly used three-dimensional curve is a straightforward extension of the two-dimensional Hilbert curve introduced by Butz \cite{BUTZpaper} and is thus identified by Haverkort \cite{haverkort} as \emph{Butz}.
Butz is one of the simpler curves with the base pattern Ca00,
which is the shape of octant traversals on each level of the curve.
With Lawder's formulation \cite{kabbalah} the curve is uniquely identified by rotations encoded by start and end points of each octant.
More generally the transformations can include mirroring the pattern so its chirality changes or inverting it such that state transformations are applied backwards.
Figure \ref{fig:3d-hilberts} shows Butz and five other curves with base pattern Ca00, and their $L_p$ dilations are in Table \ref{tab:dilations}.
The well-foldedness of curves with this pattern allow for efficient computation of the Hilbert coordinate as described in Section \ref{sec:hsfc}.
Curves with potentially better superior parameters to Butz include \emph{Alfa}, \emph{Sasburg} and \emph{Beta},
while \emph{Harmonious} and \emph{Base camp} should be worse than Butz on all metrics.
The most notable curves are the \emph{hyperorthogonal} curves Alfa and Beta, with Alfa having optimal $L_\infty$ dilation and Beta optimal $L_1$ and $L_2$ dilations.
We can additionally consider the \emph{Z-curve}; a discontinuous space-filling curve which should thus have the worst locality with effectively infinite $L_p$ dilation.

\begin{table}[!t]
    \centering
    \caption{$L_p$ dilations for evaluated curves from Haverkort \cite{haverkort}. Also shown: the Octree curve, an implementation of Butz used by the Zoltan library.}
    \begin{tabular}{l|S|S|S}
        Name & {$\sqrt[3]{\mathrm{WL}_1}$} & {$\sqrt[3]{\mathrm{WL}_2}$} & {$\sqrt[3]{\mathrm{WL}_\infty}$} \\\hline
        Butz/Octree       & 4.62 & 2.97 & 2.89 \\
        Alfa       & 4.64 & 2.84 & 2.32 \\
        Harmonious & 4.63 & 3.07 & 3.04 \\
        Sasburg    & 4.58 & 3.00 & 2.66 \\
        Base camp  & 5.27 & 3.21 & 3.04 \\
        Beta       & 4.48 & 2.65 & 2.41
    \end{tabular}
    \label{tab:dilations}
\end{table}

\section{Software} \label{sec:software}
\subsection{Vlasiator}
As a hybrid-Vlasov simulation, Vlasiator works by modeling ions as a six-dimensional distribution function propagated using the Vlasov equation.
Electrons are assumed to be a charge-neutralizing fluid and Maxwell's equations in the Darwin approximation are used to propagate electromagnetic fields;
closure is provided by MHD Ohm's law with the Hall and electron pressure gradient terms for the electric field,
and Faraday's law for the magnetic field \cite{vlasiator531,vlasiatorpaper,6dtech}.
Simulation data is on a three-dimensional orthogonal Eulerian grid,
with each cubic spatial cell containing a three-dimensional velocity grid holding the velocity distribution in that cell
as well as spatial variables calculated from the velocity moments of the distribution such as particle density and bulk velocity.

Vlasiator primarily spends computational time in spatial propagation of the distribution function.
For this, \emph{translation pencils} are formed in each direction,
where the distribution function is sampled from the Eulerian grid, reconstructed into a polynomial in Lagrangian space, advected and integrated back to the Eulerian grid \cite{vlasovmethods}.
The three-dimensional propagation is split into successive one-dimensional propagations and solved using the SLICE-3D scheme \cite{slice3d}.
This requires communicating ghost data between each step along each pencil, and is the primary communication bottleneck of the simulation.
The other significant solvers constituting the simulation propagation are velocity propagation which is local to each cell's velocity space and thus to each process,
system boundary conditions which requires communications for boundary cells only twice per timestep,
and the field solver, which is done on an uniform grid with separate, simple load balancing and thus not considered here \cite{papadakis}.

As the simulation domain is global, the resolution required is not uniform.
In regions with low spatial variance of plasma properties, a lower resolution is sufficient.
This is achieved using octree adaptive mesh refinement (AMR),
which refines and coarsens individual cells dynamically based on certain parameters \cite{amrpaper}.
Refining a cell splits into eight children up to a certain level of refinement,
and conversely eight children of the same cell can be coarsened and merged back to one.
On a production-scale run with three levels of refinement, 
the final grid size is about \SI{2.5}{\percent} compared to a uniform grid at the highest resolution.

Additionally, the vast majority of the velocity space is almost empty.
As an example, solar wind upstream from Earth is modeled flowing away from the Sun with a Maxwellian distribution,
making the value of the unperturbed distribution effectively zero where $v_x > 0$ or $\abs{v_{y,z}} \gg 0$.
To reduce computational load, some velocity space cells are pruned based on the phase space density.
The velocity space is divided into $4 \times 4 \times 4$ cell blocks and computed block by block;
a block is stored and propagated if and only if any cell inside it or in a neighboring block in six dimensions has any cell above a set density threshold.
This can cut down the size of the velocity space to under \SI{1}{\percent} of the original volume \cite{vlasiatorpaper,vlasovmethods}.
While AMR and the sparse velocity space approach massively reduce the size and computational cost of the simulation,
they also complicate load balancing.
Due to the varying velocity space block count, cells may have a difference in orders of magnitude in computational load.
Additionally, the adaptive refinement means cells can have larger amounts of spatial neighbors,
complicating the communication graph.

Spatial propagation dominates propagation time, so that is the part of simulation we want to optimize.
This consists of propagating the distribution along each dimension,
which requires data from neighboring cells along that dimension and thus ghost communication for remote cells.
In Vlasiator, spatial propagation needs to be done for each cell's active velocity space blocks;
thus for a cell's propagation weight we can simply use its block count.
Additionally, system boundary cells are assigned a weight of $1/6$ compared to other cells, as they are typically only involved in one propagation direction and are not accelerated.
The block count also corresponds to the cost of velocity propagation,
so a partition with good spatial propagation weight balance will also be well-balanced for velocity propagation.
As the block count is also the memory footprint of local (and ghost) cells
and is directly proportional to the size of required temporary propagation buffers, 
this makes memory usage between processes relatively balanced as well.


In addition to the propagation weight, PHG partitioning requires edges.
As edges correspond to communication, the relevant factor to consider is the amount of data communicated, i.e. the velocity space block count of the cell communicating.
In hypergraph partitioning, each spatial cell belongs to a hyperedge containing that cell and all its communication neighbors, with block count as the weight.
In graph partitioning, a cell has an edge for each of its neighbors, using the same weight as hypergraph partitioning.
For the communication neighborhood we use the union of spatial propagation and system boundary neighborhoods, consisting of the two nearest cells in all directions including diagonals.
Boundary cells are treated the same as other cells for edge weights, as their communication load is equal.

\subsection{Zoltan}
Vlasiator uses Zoltan as its load balancing library \cite{zoltan},
which implements every algorithm mentioned in Section \ref{sec:algs} natively, using PHG for both graph and hypergraph partitioning.
For the HSFC method, Zoltan's native \emph{Octree} curve is an implementation of Butz.
We implemented Zoltan-compatible state-tables for the six curves discussed in Section \ref{sec:curve-optimization} by adapting the algorithm provided by Lawder and King \cite{kabbalah}, in addition to a re-implementation of the Butz curve with different coordinate order.
Additionally we considered the three-dimensional Z-curve which should have comparatively worse locality due to discontinuities.

\section{Methodology} \label{sec:methods}
We tested load balancing performance on two different-scale Vlasiator runs;
a smaller standard sample run referred to as S and an adapted production-scale run referred to as L.
The purpose was to measure the scalability of different methods,
as well as evaluate performance in both simple and complex grid configurations.
The sample run S has a spatial grid with a base size of $51 \times 40 \times 40$ cells with three levels of AMR,
and velocity space extents of up to \num{200} cells in each direction.
Meanwhile L has a spatial grid of $70 \times 50 \times 50$ cells with the same amount of AMR levels,
and up to \num{280} velocity space cells in each direction.
The S run covers \num{200} timesteps from initialization of the simulation, 
while L runs for \num{50} timesteps starting from an existing checkpoint file for a more representative sample of a long simulation.
In run S load balancing was done at a static frequency of every \num{50} timesteps including initialization,
while on run L every \num{10} timesteps including initialization.
Table \ref{tab:runs-cellcounts} contains the amount of spatial and velocity space cells of the two runs,
as well as the percentage of cells on each refinement level.
Run L is roughly two orders of magnitude larger than S in both the amount of spatial cells and velocity space cells.
In addition to the larger size of the unrefined grid, L has much more aggressive AMR which can be seen in the relatively low count of level 0 and 1 cells.

\begin{table*}[!t]
    \centering
    \caption{
        Spatial and velocity space cell counts of the two runs,
        along with the percentage of cells on each refinement level.
    }
    \begin{tabular}{c|c|c|c|c|c|c}
        &&&\multicolumn{4}{c}{Spatial cells on refinement level}    \\
        Run & {Velocity space cells} & {Spatial cells} & {L0} & {L1} & {L2} & {L3} \\\hline
        S & \num{4944761216} & \num{285972} & \SI{26.4}{\percent} & \SI{12.3}{\percent} & \SI{36.5}{\percent} & \SI{24.8}{\percent} \\
        L & \num{709520319936} & \num{11940800} & \SI{6.26}{\percent} & \SI{6.85}{\percent} & \SI{30.6}{\percent} & \SI{56.3}{\percent}
    \end{tabular}
    \label{tab:runs-cellcounts}
\end{table*}

For both test set-ups, 
we ran 12 jobs on the EuroHPC LUMI supercomputer with identical physical parameters and Slurm configuration.
We compared all six curves described in Section \ref{sec:curve-optimization} as well as Zoltan's basic Octree HSFC implementation and the Z-curve,
as well as all five methods described in Section \ref{sec:algs}, using the best curve identified for HSFC which was Beta (Section \ref{sec:comparison-hilberts}) and PHG for graph and hypergraph partitioning.
Each of the twelve parallel jobs ran every curve and load balancing method in succession to mitigate variance in node performance and inter-node communication between tests.
For S we used 8 nodes with 16 tasks on each node and 16 threads on each task, and for L 500 nodes with the same task and thread counts on each nodes.
The CPU nodes used have two AMD EPYC 7763 CPUs with 64 cores each for 128 physical cores and 256 logical cores with hyperthreading, and 256 GiB of memory \cite{lumihw}.

The run times were compared using the profiling library Phiprof, which provides timers on relevant sections of the simulation code \cite{phiprof}.
The timers inspected are total simulation propagation time (Propagate), spatial propagation (Spatial-space), velocity space propagation (Velocity-space) and system boundary updates (System-boundaries).
Note that total propagation is not a sum of the other timers as some timer regions are not included,
most notably the field solver which isn't affected by load balancing.

The simulation data analyzed as well as run configurations and analysis scripts are provided in Fairdata Qvain \cite{config-files}.
Analysis is done by parsing plaintext Phiprof output as well as reading simulation data using the Analysator library \cite{analysator}.

\section{Results} \label{sec:results}
\subsection{Comparison of algorithms} \label{sec:comparison-algs}
\begin{figure*}[!t]
    \centering
    \subfloat[]{
        \includegraphics[width=3in]{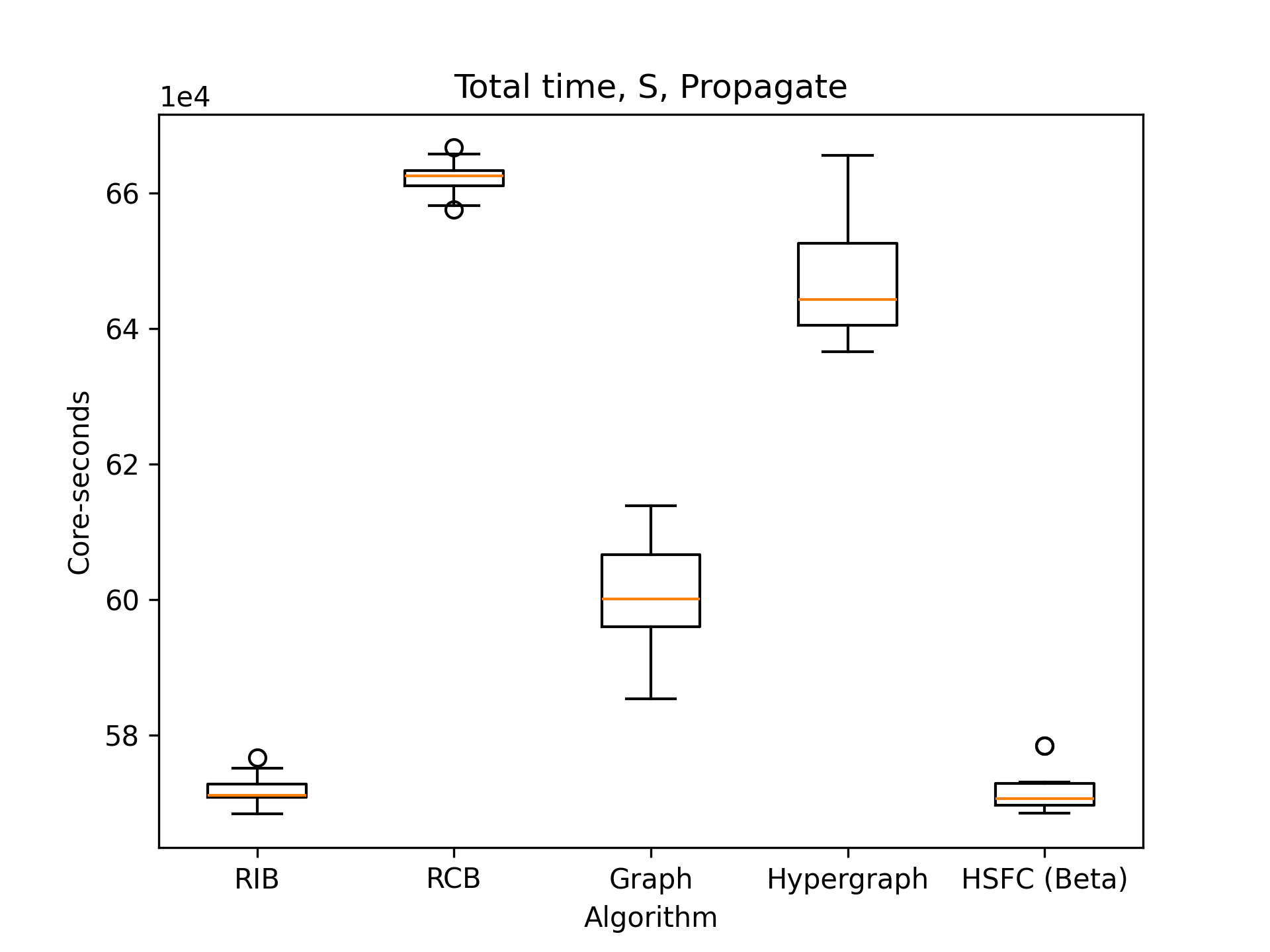}
    } \hfil
    \subfloat[]{
        \includegraphics[width=3in]{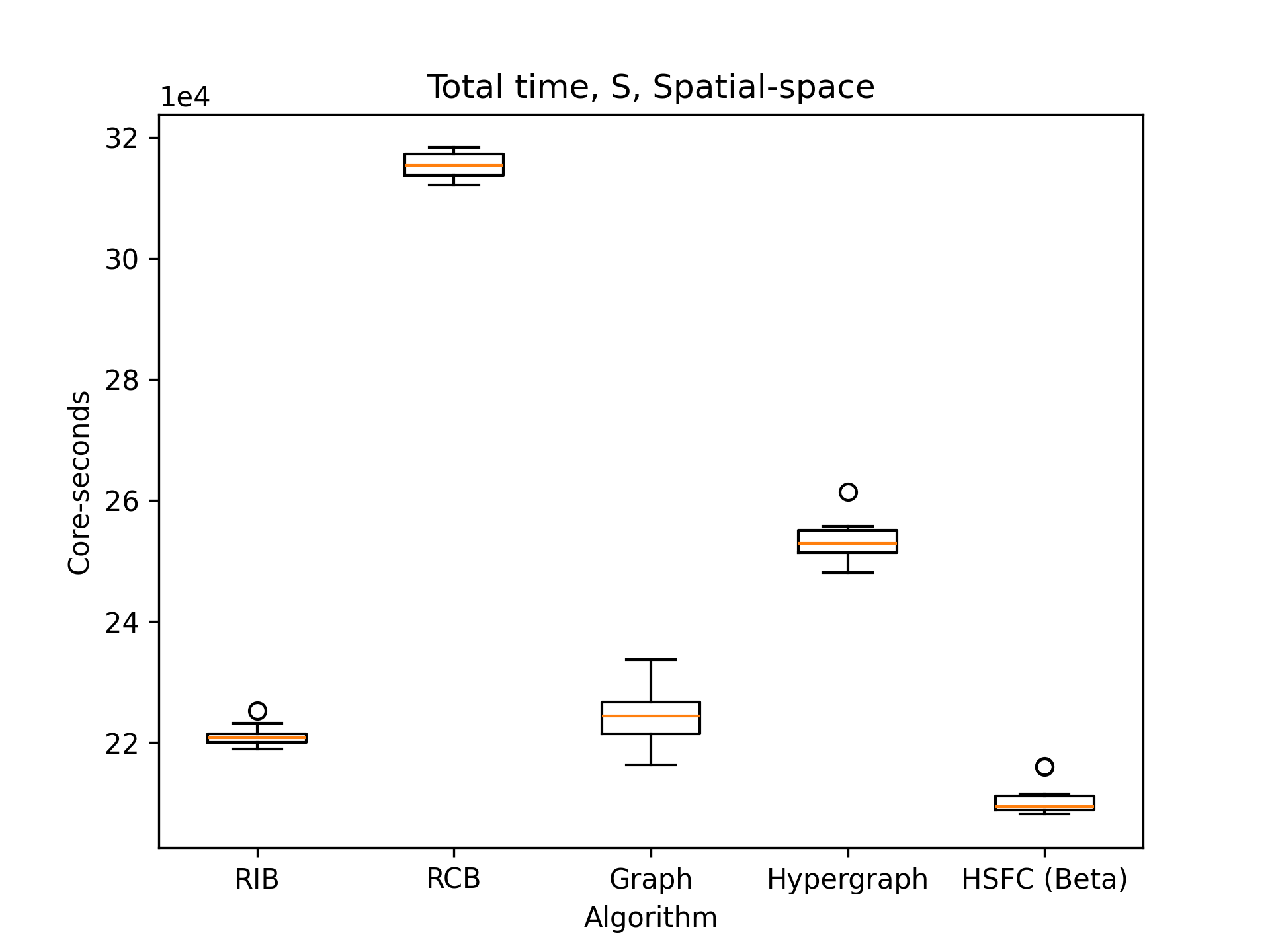}
    } \\
    \subfloat[]{
        \includegraphics[width=3in]{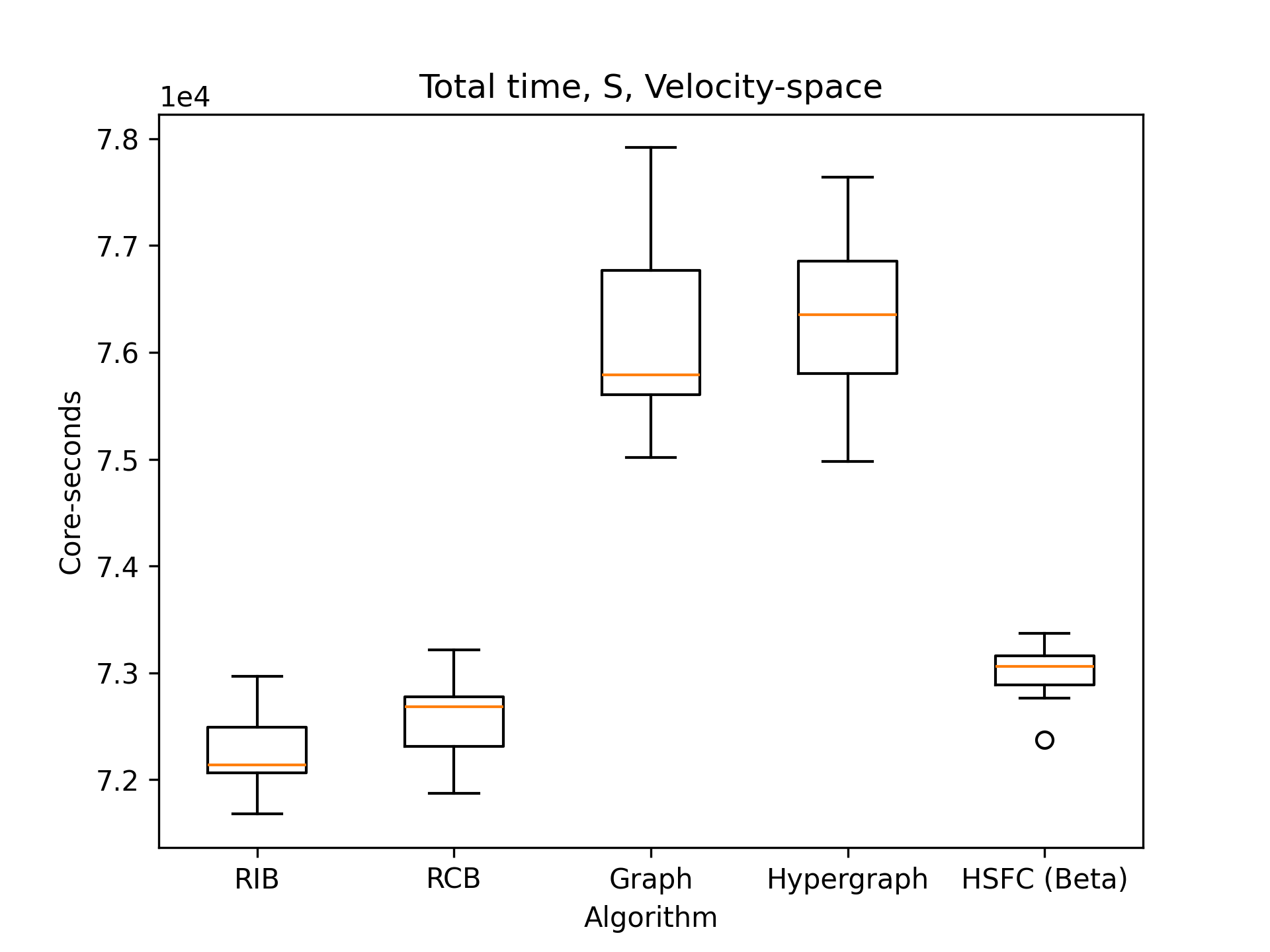}
    } \hfil
    \subfloat[]{
        \includegraphics[width=3in]{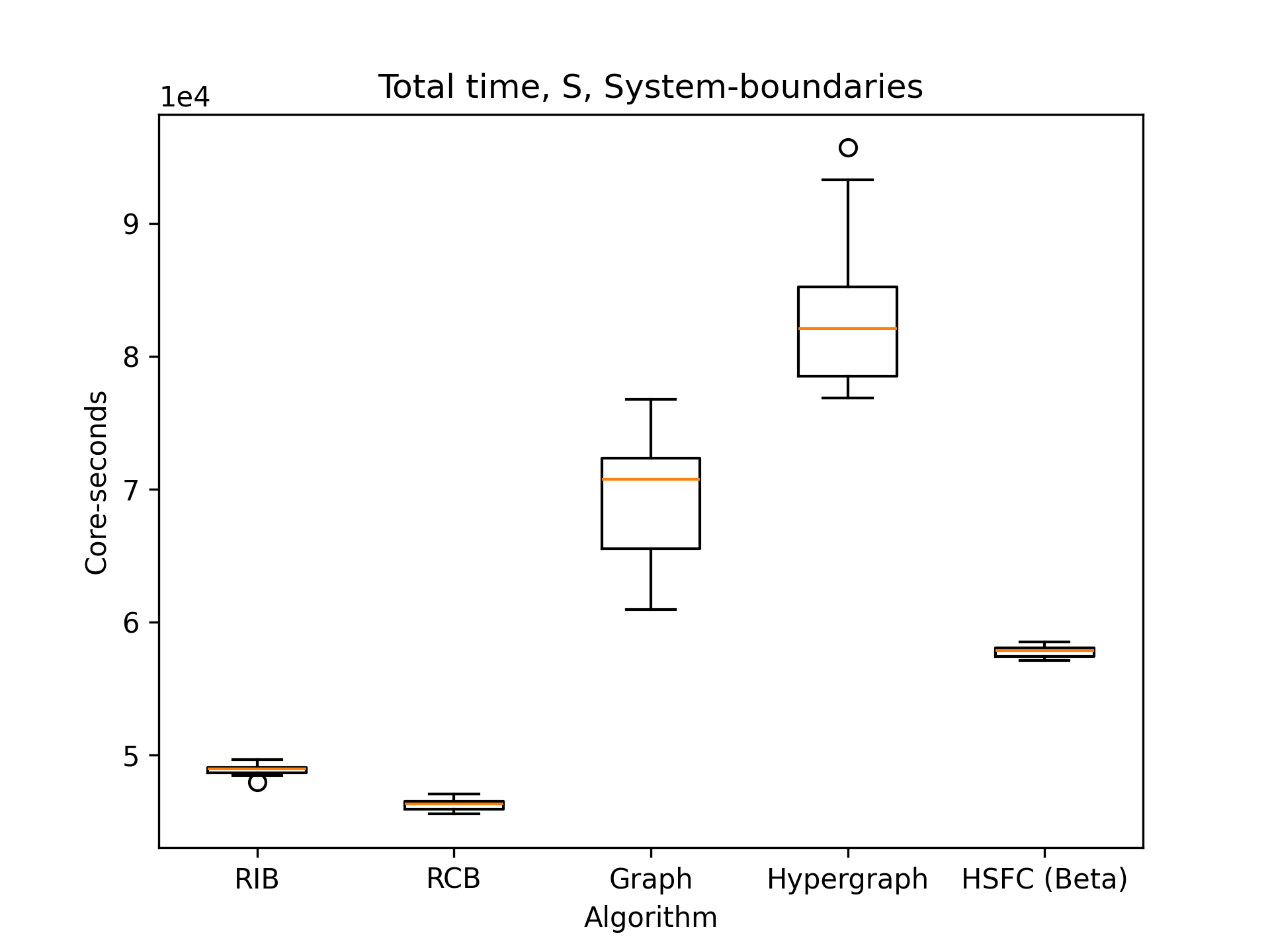}
    }
    \caption{
        Box plots of core-seconds spent in total propagation (a), spatial propagation (b), velocity space propagation (c) and system boundary updates (d) for algorithms tested on 12 trials of run S, where the Beta curve is used for HSFC.
        The box extends from the first to the third quartile of the measurements with a horizontal line at the median, with whiskers showing the furthest data points within \num{1.5} times the inter-quartile range, and the circles showing outliers outside the whiskers.
        Propagation time is dominated by spatial space, where HSFC performs the best. However, worse performance in velocity space and system boundary updates bring total propagation time in line with RIB.
    }
    \label{fig:sample-timers-algs}
\end{figure*}

\begin{figure*}[!t]
    \centering
    \subfloat[]{
    \includegraphics[width=3in]{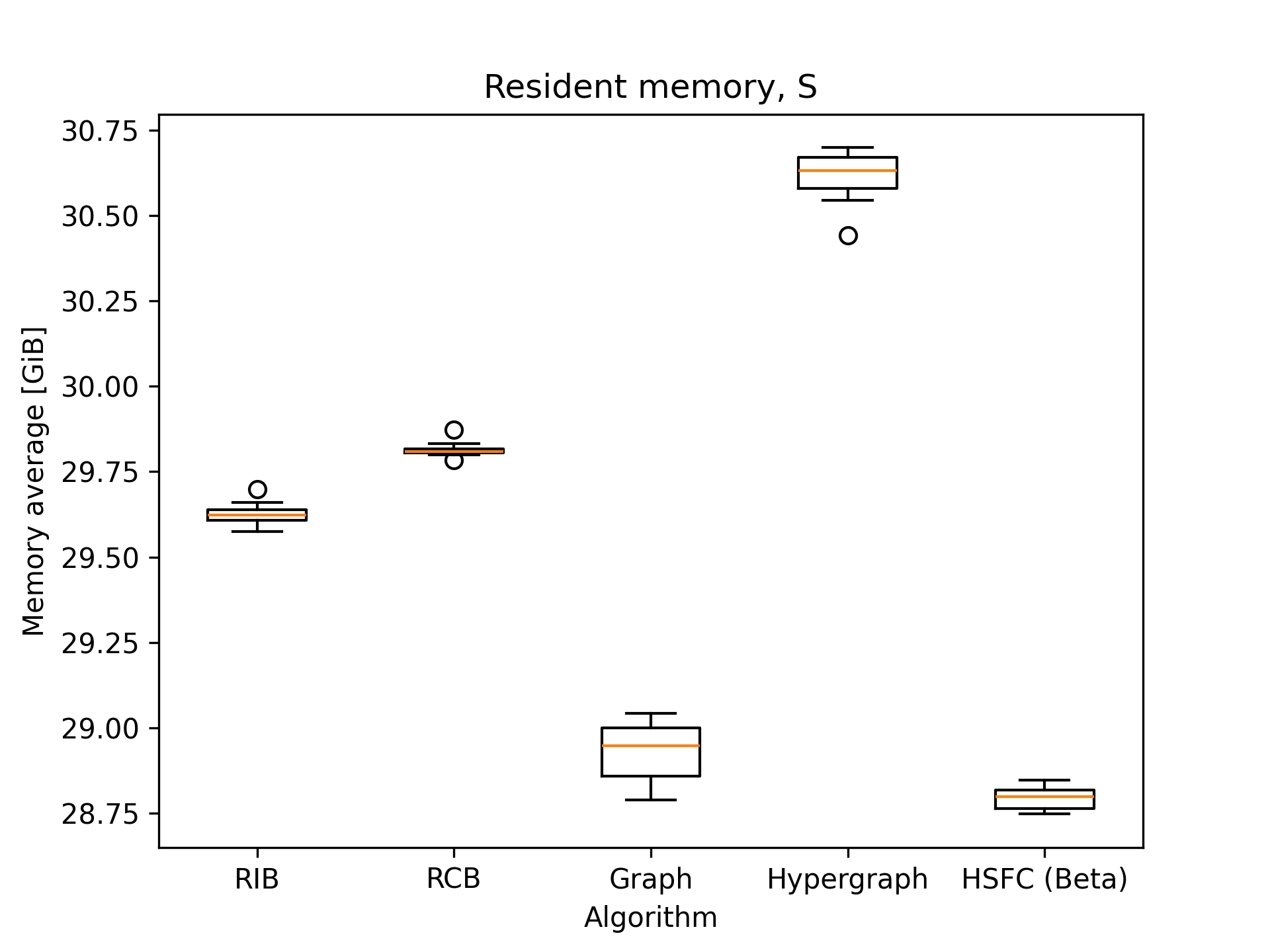}
    } \hfil
    \subfloat[]{
        \includegraphics[width=3in]{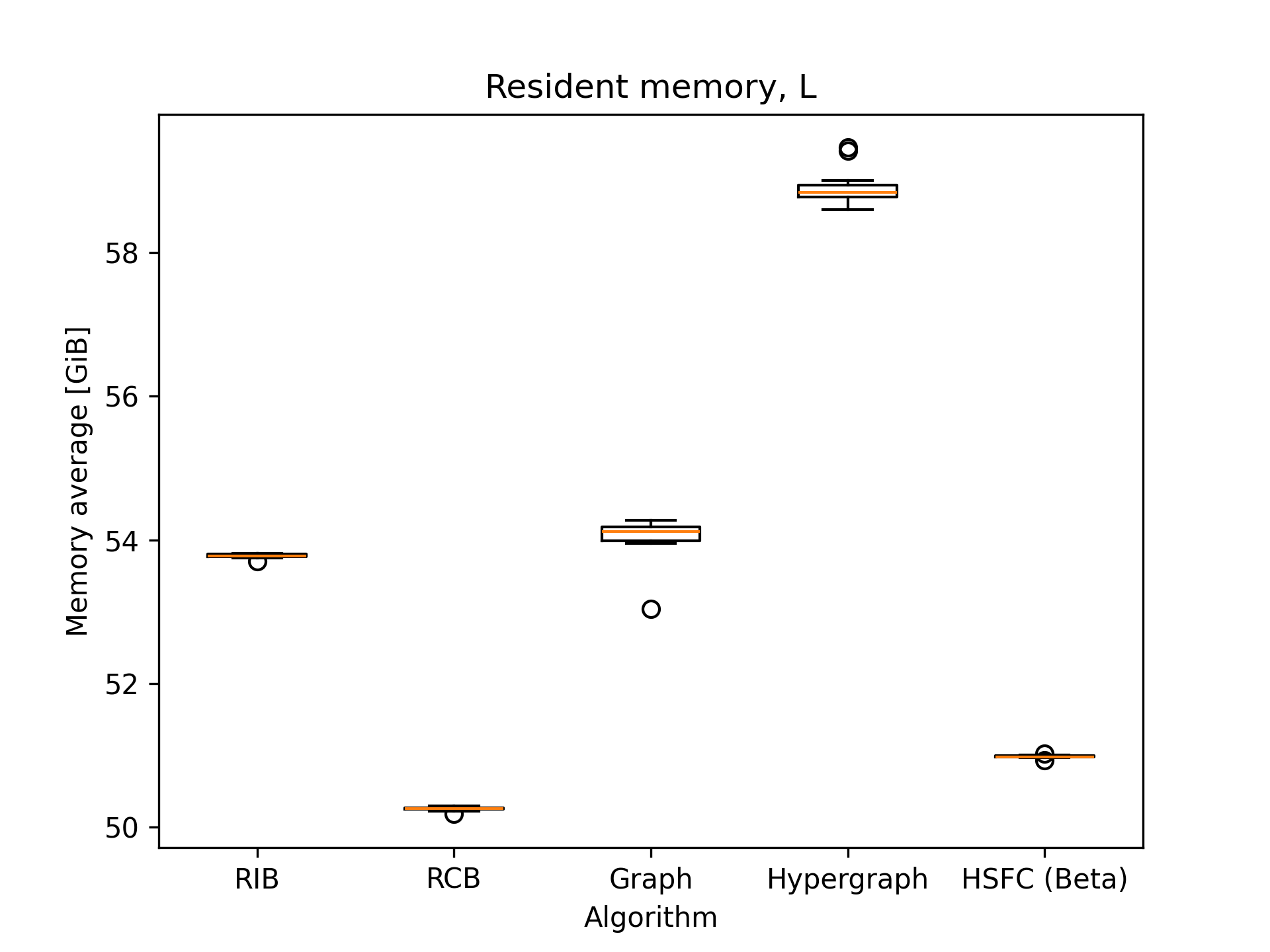}
    }
    \caption{
        Box plots of resident memory per node at the end of simulation for the algorithms tested on run S (a) and run L (b), 12 trials.
        Box plot parameters as in Figure \ref{fig:sample-timers-algs}.
    }
    \label{fig:mem-algs}
\end{figure*}

We first consider the performance of the different algorithms.
The timers for the solver sections and five algorithms considered for run S are shown in Figure \ref{fig:sample-timers-algs}.
The best performing algorithms overall were RIB and HSFC, followed by Graph, Hypergraph and RCB in descending order.
The performance of RIB and HSFC was almost identical, with the difference being more in HSFC's favor in spatial propagation with a difference of about $5\,\%$.
RIB and RCB performed better in velocity space and system boundary updates, 
which implies a better balance in cell weights compared to the other algorithms.
Figure \ref{fig:mem-algs} shows the resident memory usage of different algorithms,
with run S in panel (a).
Memory usage was highest with Hypergraph, while Graph and HSFC had somewhat lower memory usage than RIB,
implying smaller ghost domains.

\begin{figure*}[!t]
    \centering
    \subfloat[]{
        \includegraphics[width=3in]{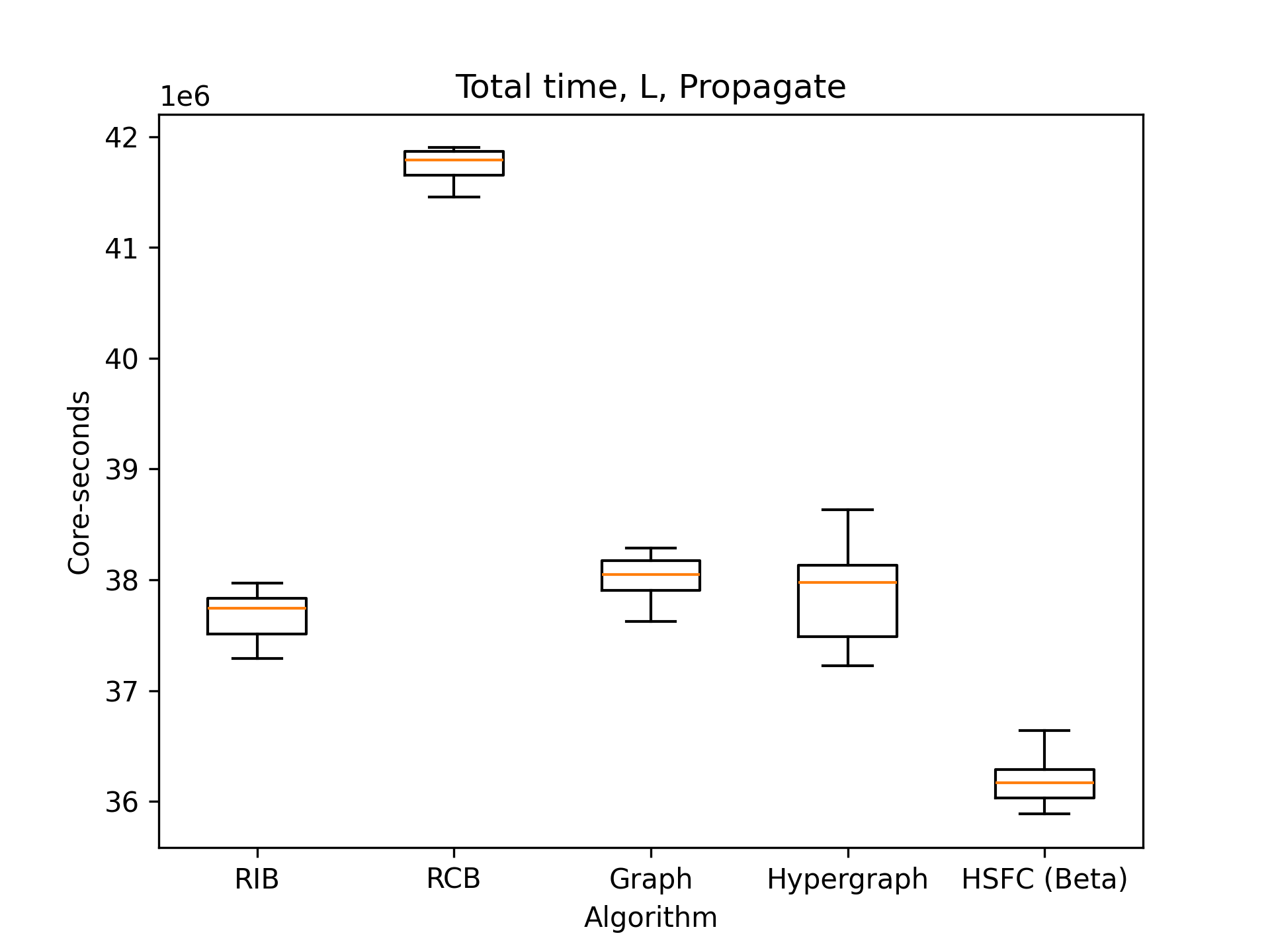}
    } \hfil
    \subfloat[]{
        \includegraphics[width=3in]{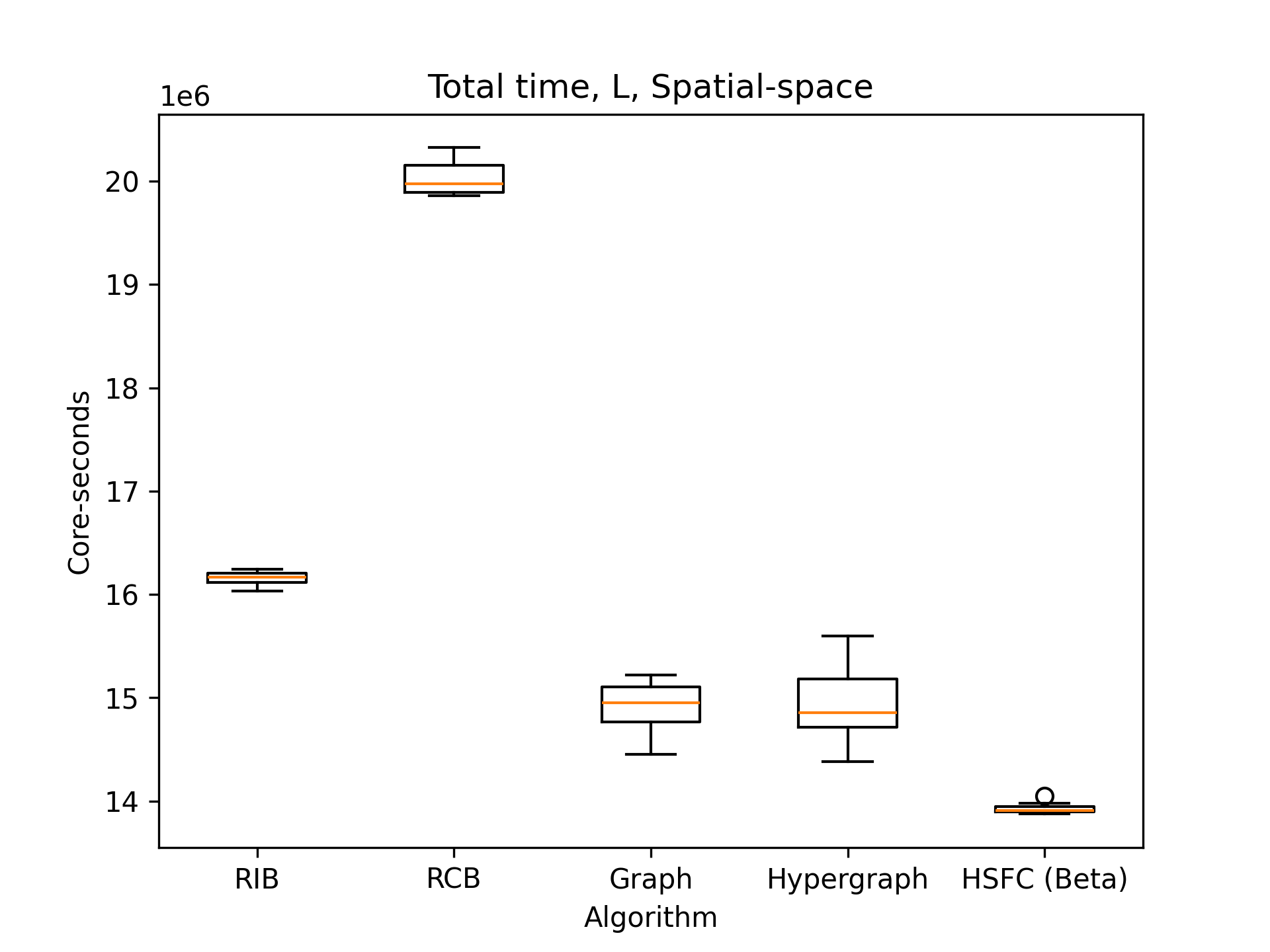}
    } \\
    \subfloat[]{
        \includegraphics[width=3in]{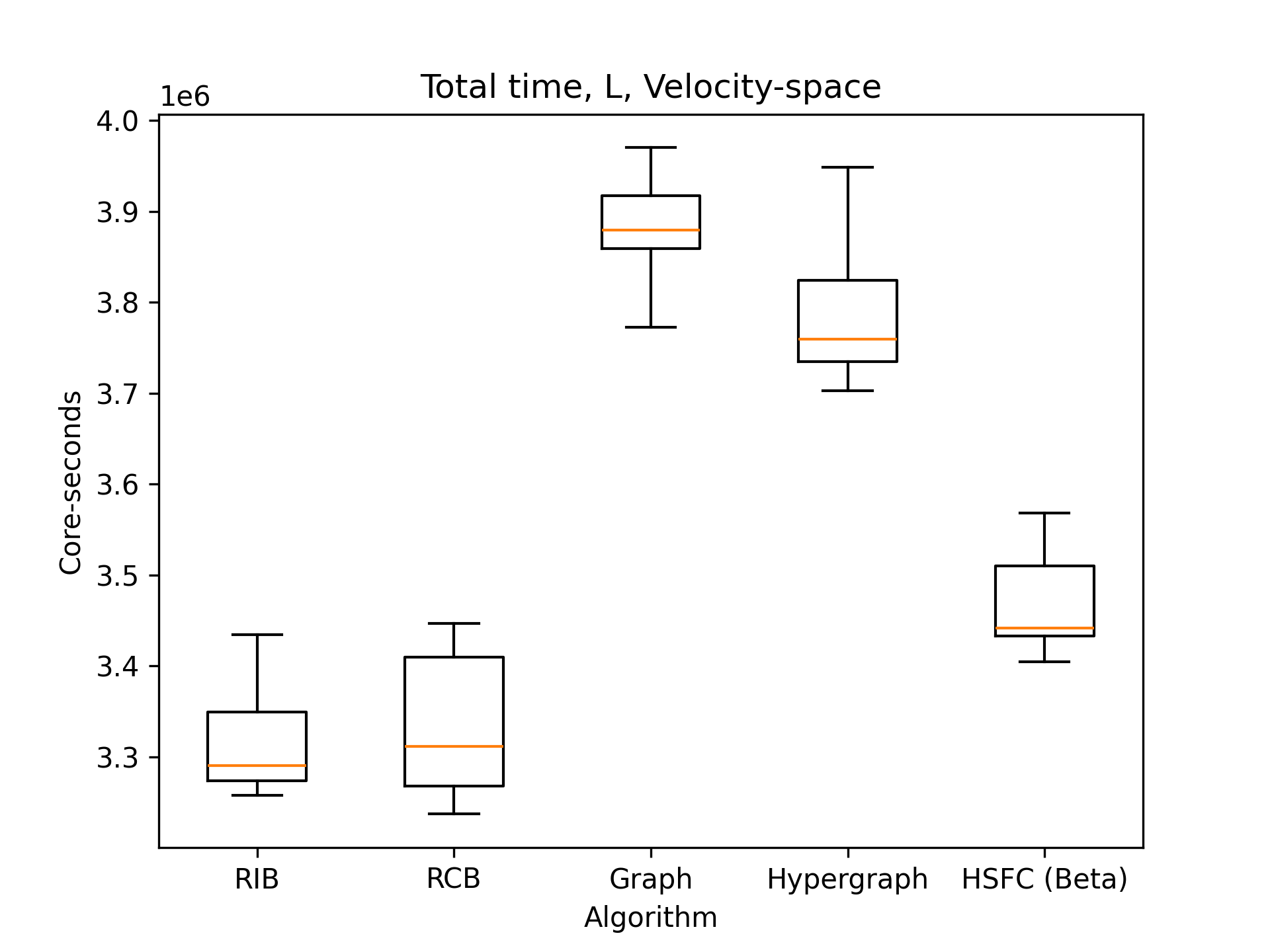}
    } \hfil
    \subfloat[]{
        \includegraphics[width=3in]{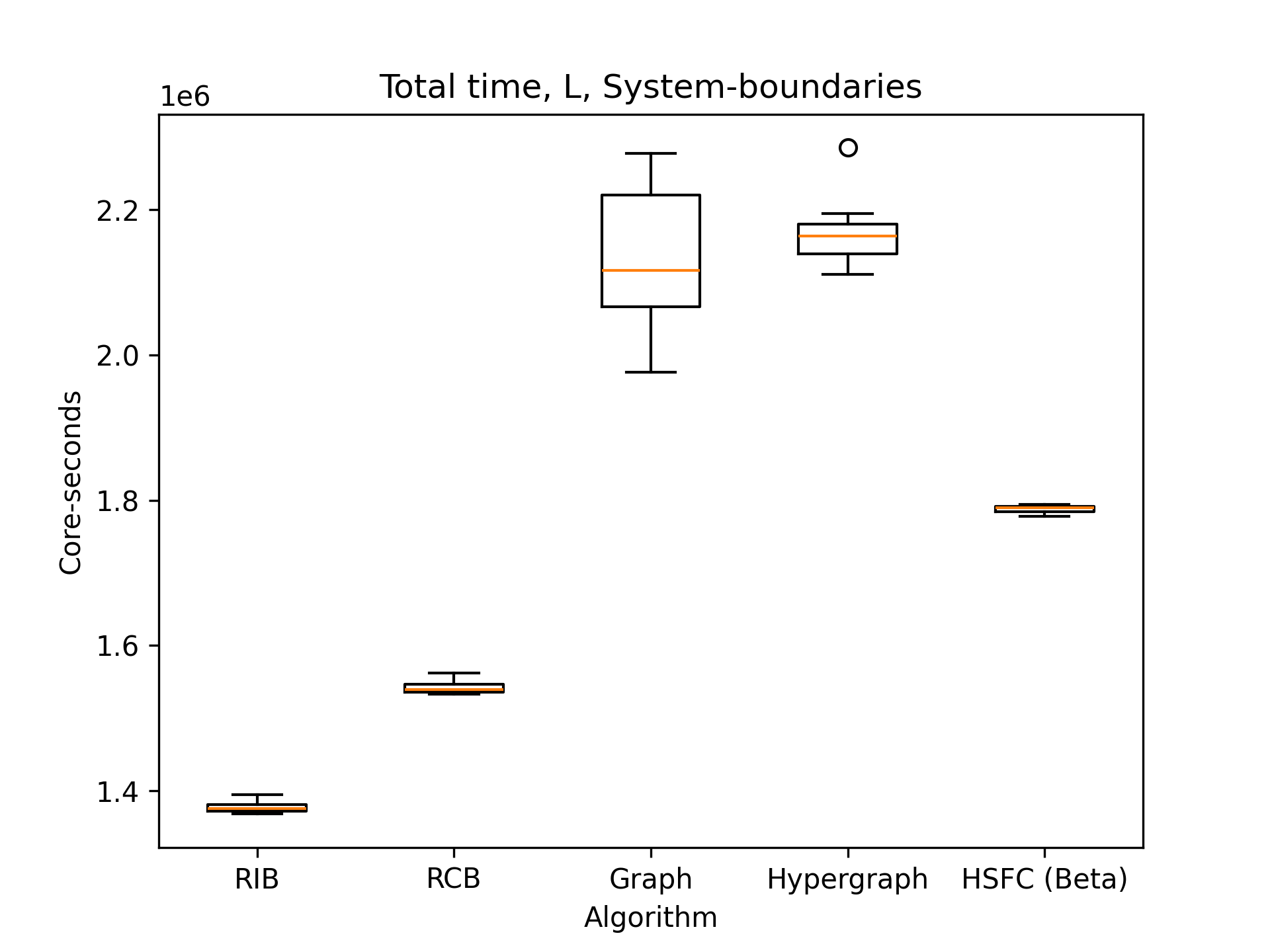}
    }
    \caption{
        Box plots of core-seconds spent in total propagation (a), spatial propagation (b), velocity space propagation (c) and system boundary updates (d) for algorithms tested on 12 trials of run L, with Beta used for HSFC.
        Box plot parameters as in Figure \ref{fig:sample-timers-algs}.
        HSFC performs the best in both spatial space and total propagation time despite the worse performance in velocity space and system boundaries.
        Graph and hypergraph partitioning also perform better than RIB in spatial space, but their worse performance in other categories bring them in line with RIB.
    }
    \label{fig:fic-timers-algs}
\end{figure*}

The timers for the solver sections and five algorithms considered for run L are shown in Figure \ref{fig:fic-timers-algs}.
The differences were more pronounced than on run S;
Graph and Hypergraph had roughly similar performance to RIB, 
with HSFC performing better by about $4\,\%$ and RCB significantly worse.
In spatial propagation the difference between HSFC and RIB is about $14\,\%$ in favor of HSFC.
Graph and Hypergraph partitioning also outperformed RIB in spatial propagation;
the reason this doesn't result in better performance overall is the worse performance in velocity space and system boundary updates.
RIB and RCB are again the best outside spatial propagation,
but HSFC's superior spatial performance makes it the best option overall.
In terms of memory usage shown in Figure \ref{fig:mem-algs} (b), 
RCB and HSFC performed the best, with Hypergraph partitioning having significantly higher memory usage than RIB.

We additionally compared the computational weight imbalances $\epsilon$ for all algorithms:
\begin{equation}
    \epsilon = \frac{\mathrm{max} \qty{{w_i} : i \in [0, n)}}{W / n} - 1, \label{eq:epsilon}
\end{equation}
where $w_i$ is the total computational weight of process $i$, $n$ is the number of processes and $W = \sum_0^{n-1} w_i$ is the total computational weight of all processes. 
This is equivalent to the approximation ratio from Section \ref{sec:theory}, substituting ideal load balance for the unknown optimal solution.
Load imbalance was determined from the first timestep of a single run with each load balancing method and is listed in Table \ref{tab:weights_algorithms}. As expected, the imbalance correlates well with time spent in velocity and system boundary updates; 
RCB and RIB have the lowest weight imbalances followed by HSFC, and the PHG methods are larger by \numrange{1}{2} orders of magnitude.
Note as well that run L had roughly double the imbalance compared to run S for all algorithms;
the larger run is harder to balance due to higher problem size in both cell count and process count, as well as a more inhomogeneous state.
Despite this, both HSFC and PHG methods outperform RIB and RCB spatial propagation on run L.
This is a result of optimizing communication, which has no effect on the other timers.

\begin{table}[!t]
    \centering
    \caption{
        Weight imbalances $\epsilon$ of the different algorithms for both runs S and L, as defined in Equation \eqref{eq:epsilon}.
    }
    \begin{tabular}{l|c|c}
         Name        & {$\epsilon$ (S)} & {$\epsilon$ (L)} \\\hline
         RCB         & \num{1.01E-3} & \num{1.82E-3} \\
         RIB         & \num{1.01E-3} & \num{2.08E-3} \\
         Graph       & \num{5.04E-2} & \num{1.00E-1} \\
         Hypergraph  & \num{5.06E-2} & \num{1.00E-1} \\
         HSFC (Beta) & \num{1.20E-3} & \num{2.30E-3}
    \end{tabular}
    \label{tab:weights_algorithms}
\end{table}


%

\subsection{Comparison of Hilbert curves} \label{sec:comparison-hilberts}
\begin{figure*}[!t]
    \centering
    \subfloat[]{
        \includegraphics[width=3in]{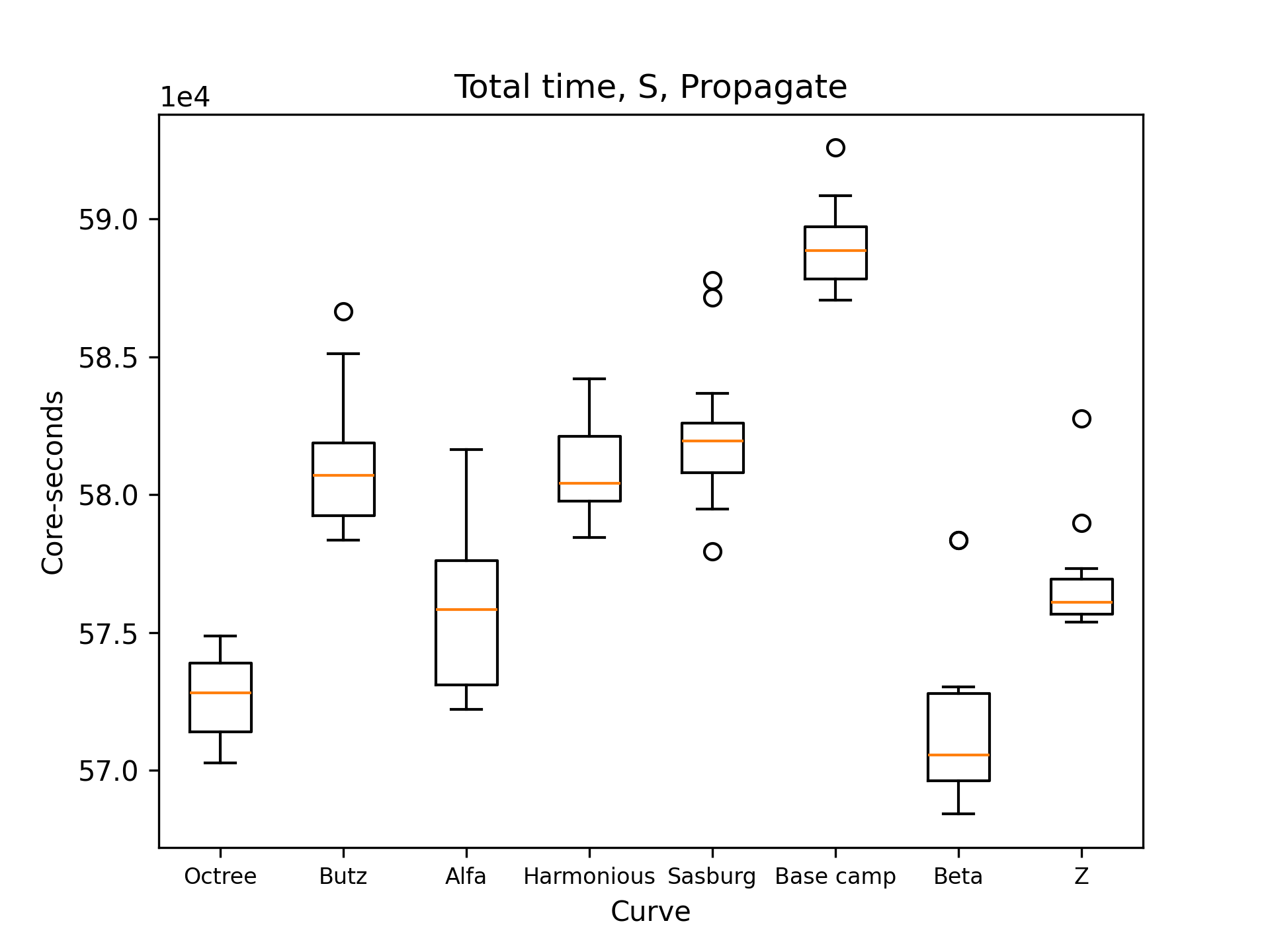}
    } \hfil
    \subfloat[] {
        \includegraphics[width=3in]{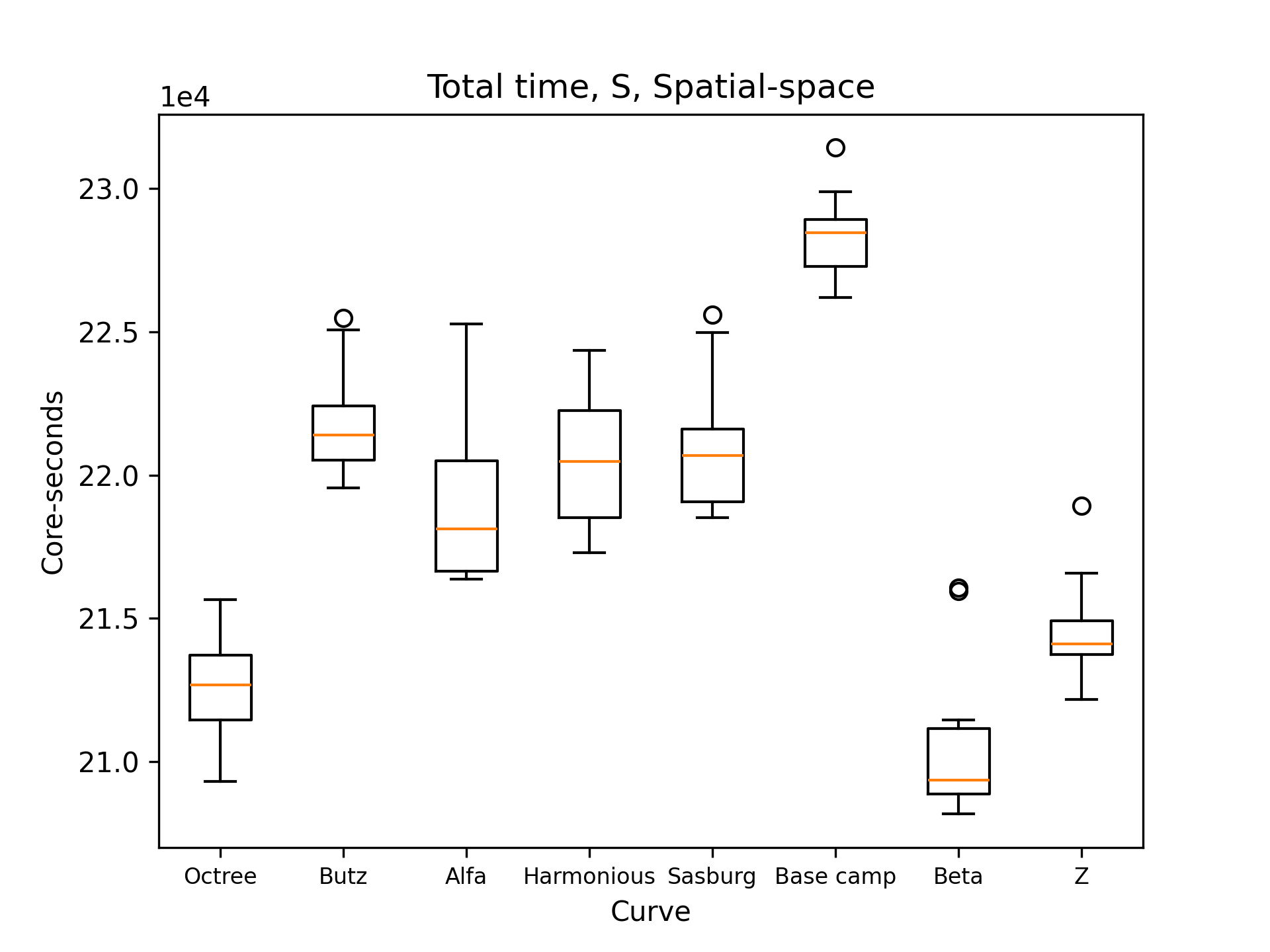}
    } \\
    \subfloat[] {
        \includegraphics[width=3in]{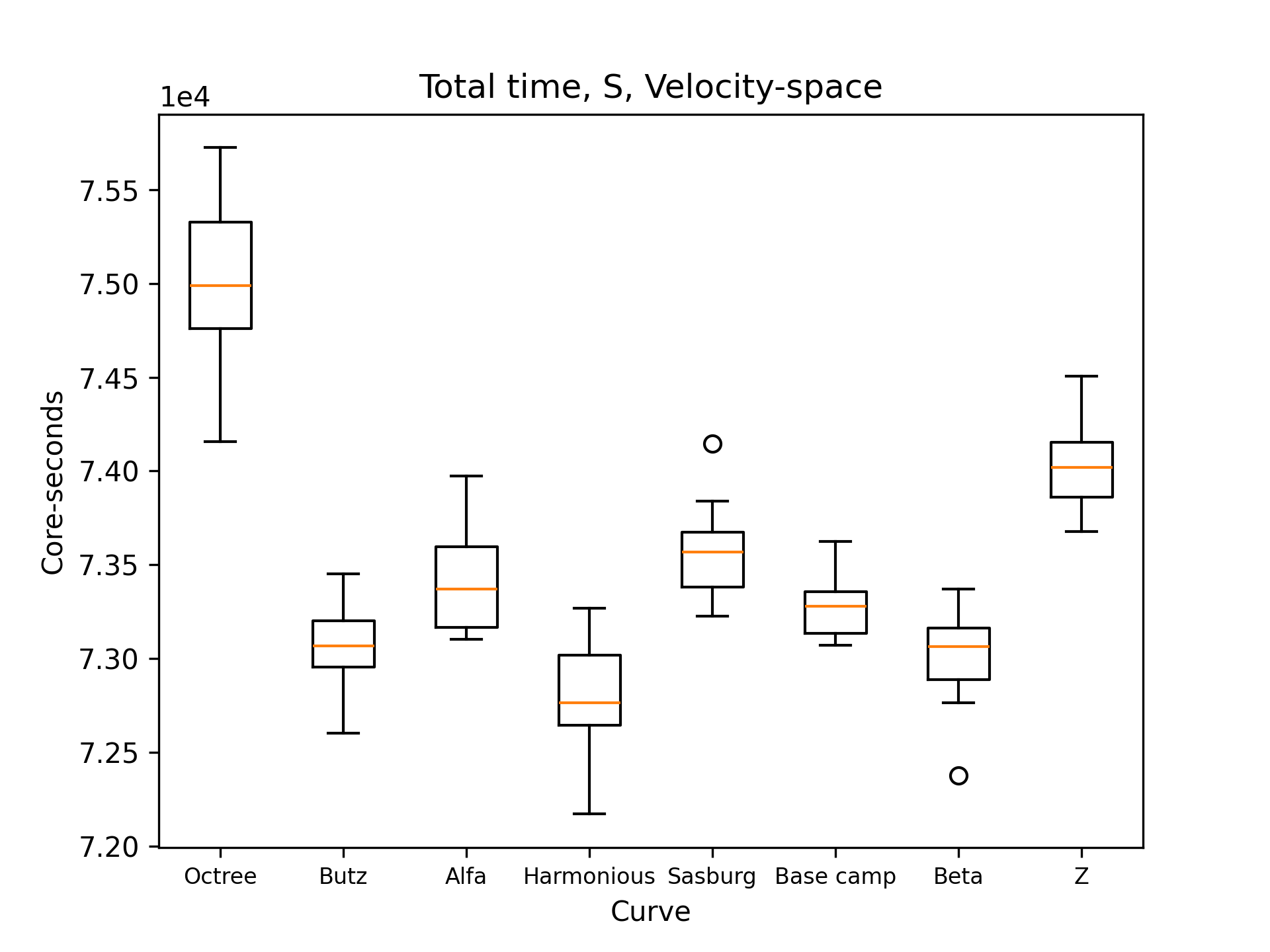}
    } \hfil
    \subfloat[] {
        \includegraphics[width=3in]{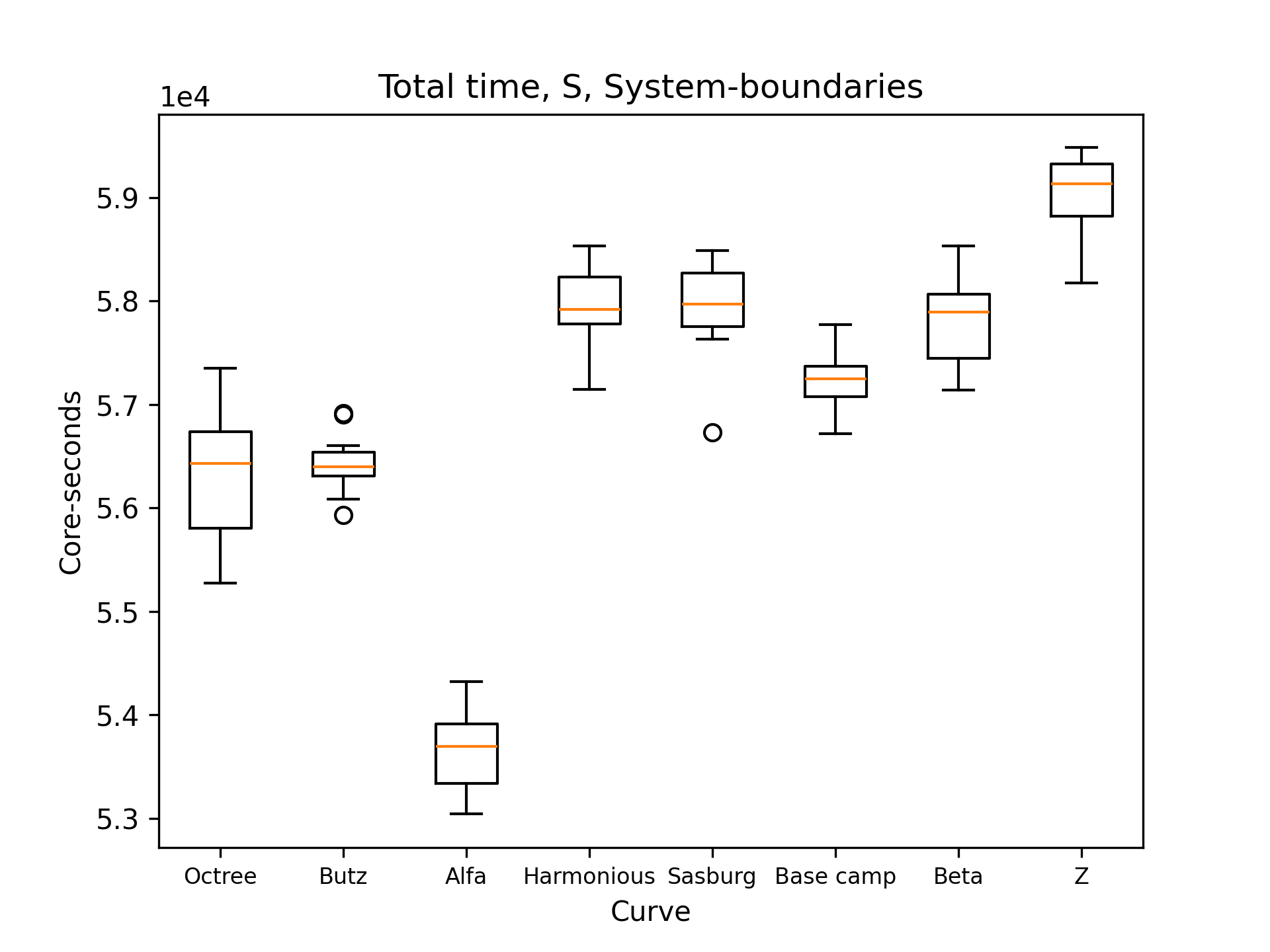}
    }
    \caption{
        Box plots of core-seconds spent in total propagation (a), spatial propagation (b), velocity space propagation (c) and system boundary updates (d) for the space filling curves tested on 12 trials of run S.
        Box plot parameters as in Figure \ref{fig:sample-timers-algs}.
    }
    \label{fig:sample-timers-hilb}
\end{figure*}

\begin{figure*}[!t]
    \centering
    \subfloat[]{
        \includegraphics[width=3in]{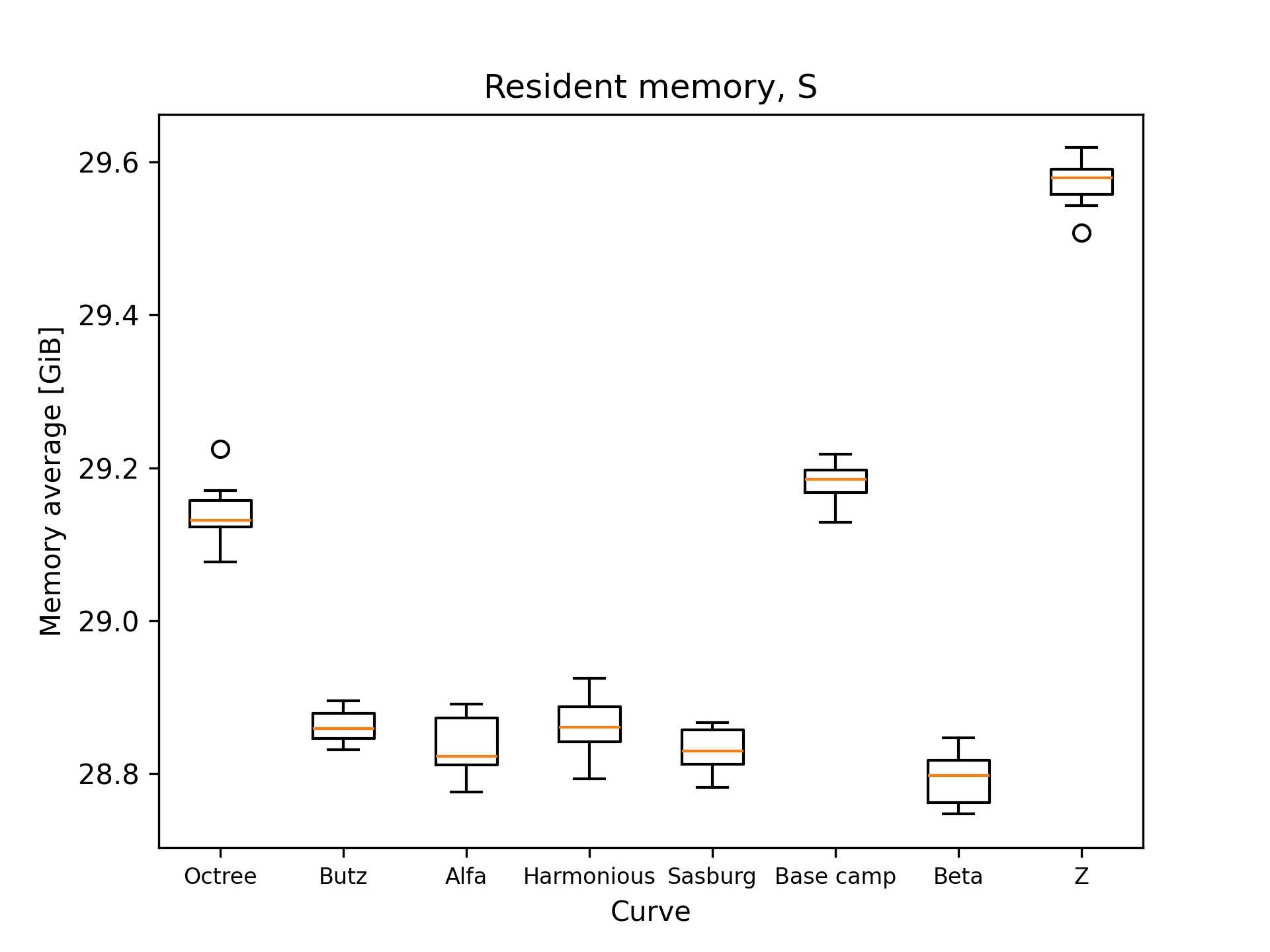}
    } \hfil
    \subfloat[]{
        \includegraphics[width=3in]{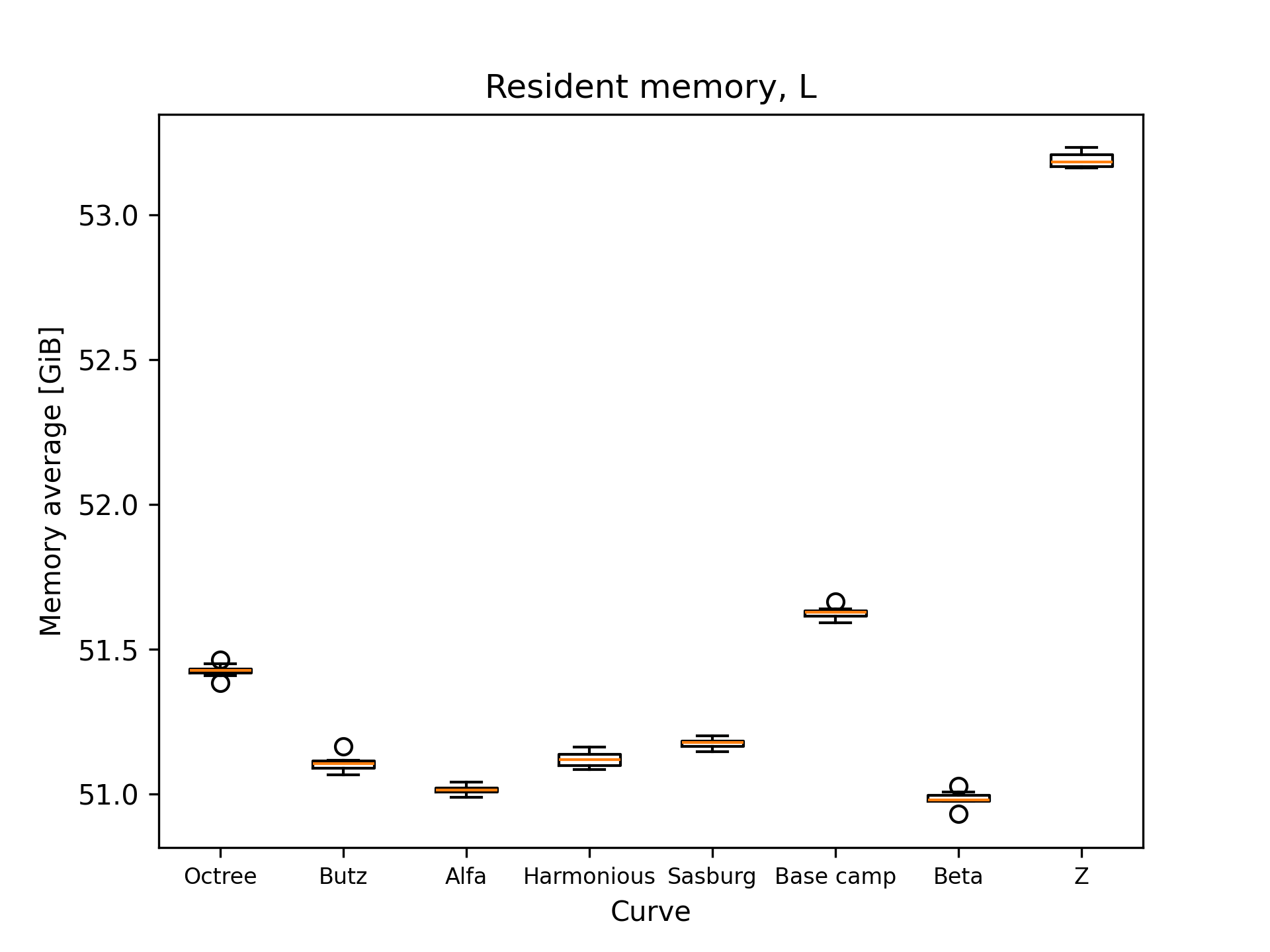}
    }
    \caption{
        Box plot of resident memory for the space filling curves tested on runs S (a) and L (b), 12 trials. The poor locality of the Base camp and Z-curves can be seen in the higher memory usage, as average memory per process corresponds to higher ghost cell count.
        Box plot parameters as in Figure \ref{fig:sample-timers-algs}.
    }
    \label{fig:mem-hilb}
\end{figure*}

As we established the HSFC method to perform the best in Section \ref{sec:comparison-algs},
let's now examine the performances of the different space-filling curves implemented.
The timers for the solver sections and eight curves considered for run S are shown in Figure \ref{fig:sample-timers-hilb}.
Most curves showed no improvement over Octree,
with Base camp being the worst.
However, Beta performed slightly better;
total propagation time is almost identical between Beta and Octree, but differences can be seen in spatial propagation,
where Beta is about $1\,\%$ faster than Octree.
Figure \ref{fig:mem-hilb} shows the resident memory usage of the different curves for both runs,
with run S in panel (a) and run L in panel (b).
There is little difference between the curves in memory overhead in run S,
but Z-curve has the highest usage, likely due to it being discontinuous.

\begin{figure*}[!t]
    \centering
    \subfloat[]{
        \includegraphics[width=3in]{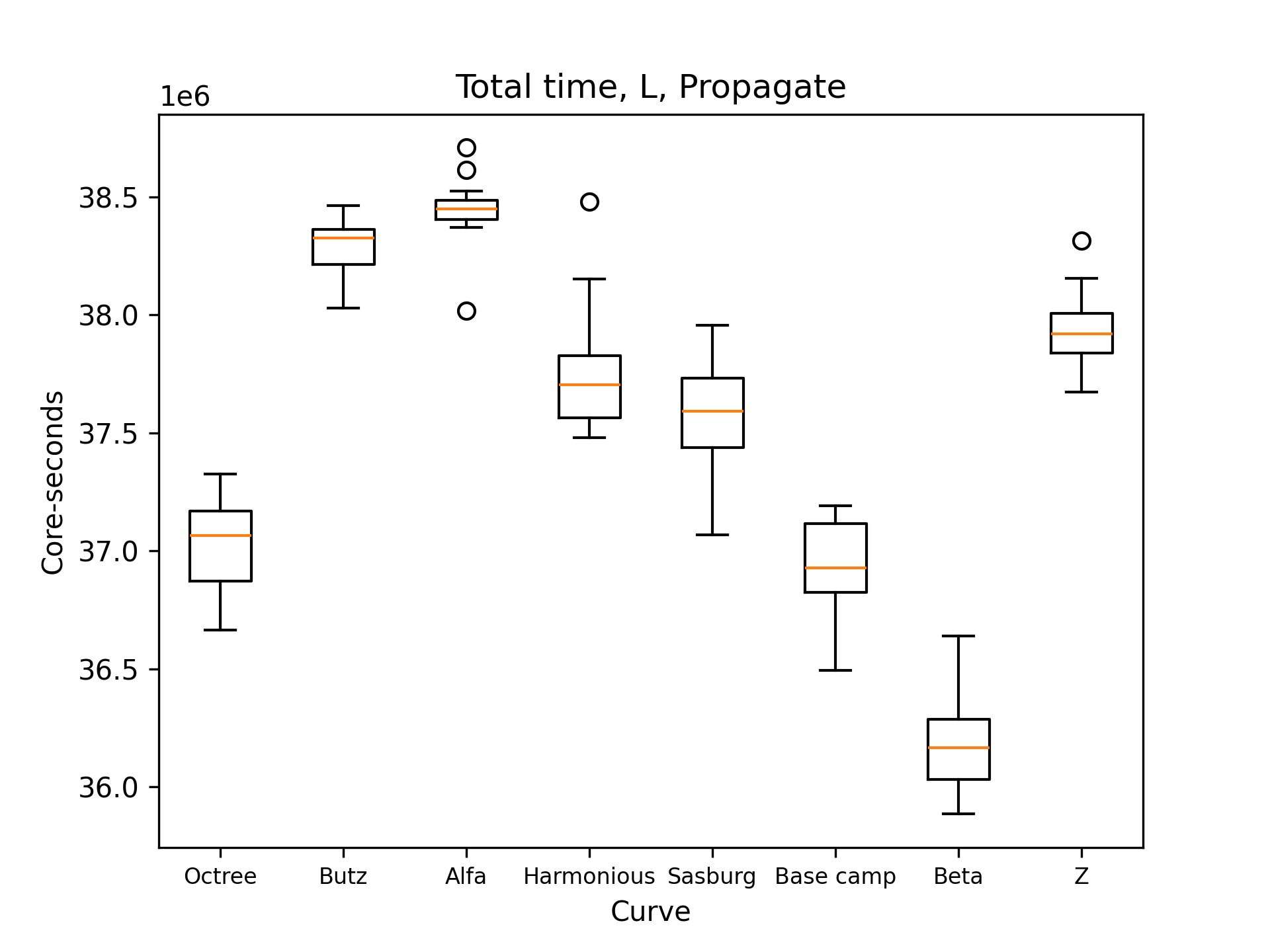}
    } \hfil
    \subfloat[]{
        \includegraphics[width=3in]{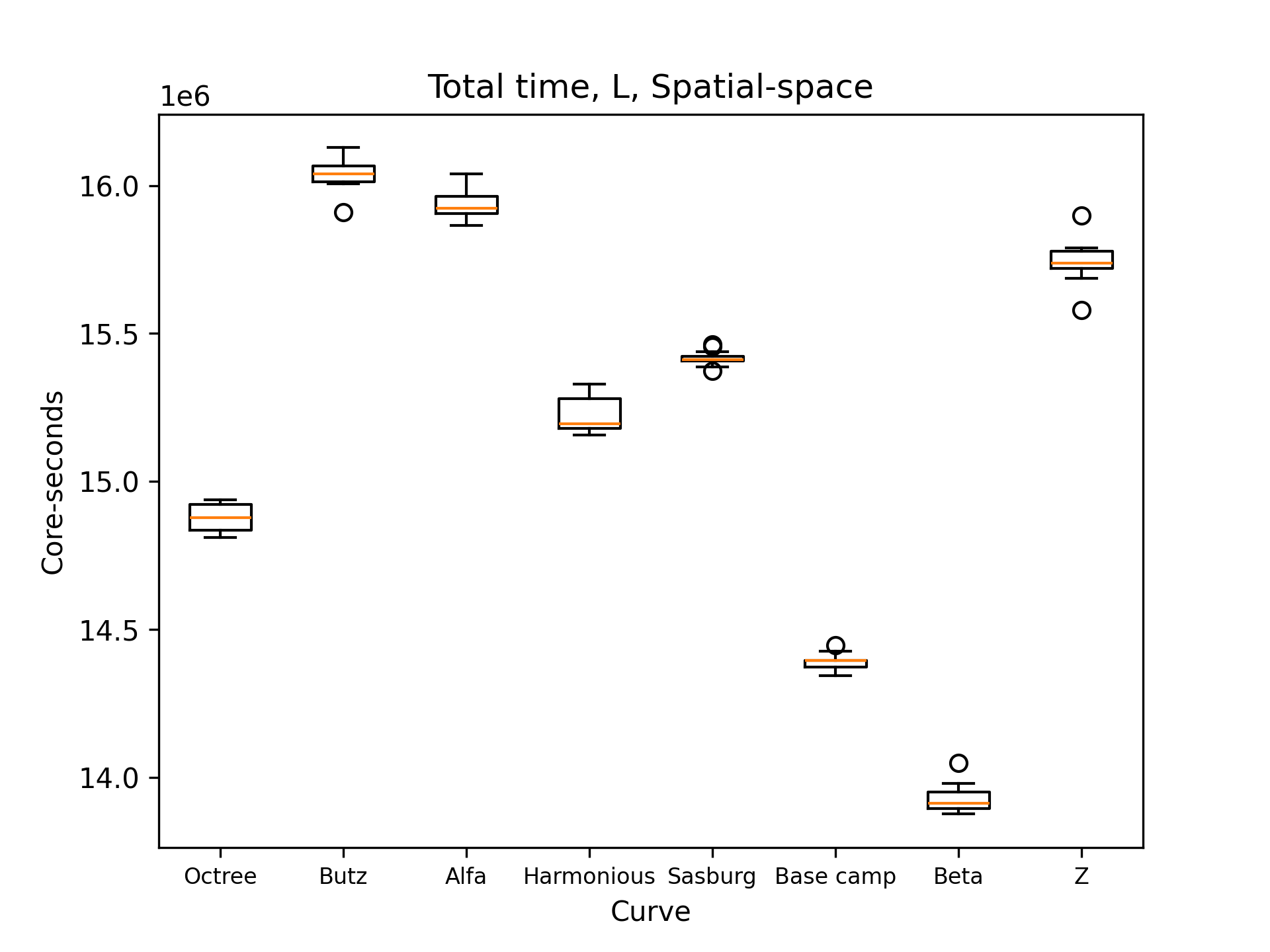}
    } \\
    \subfloat[]{
        \includegraphics[width=3in]{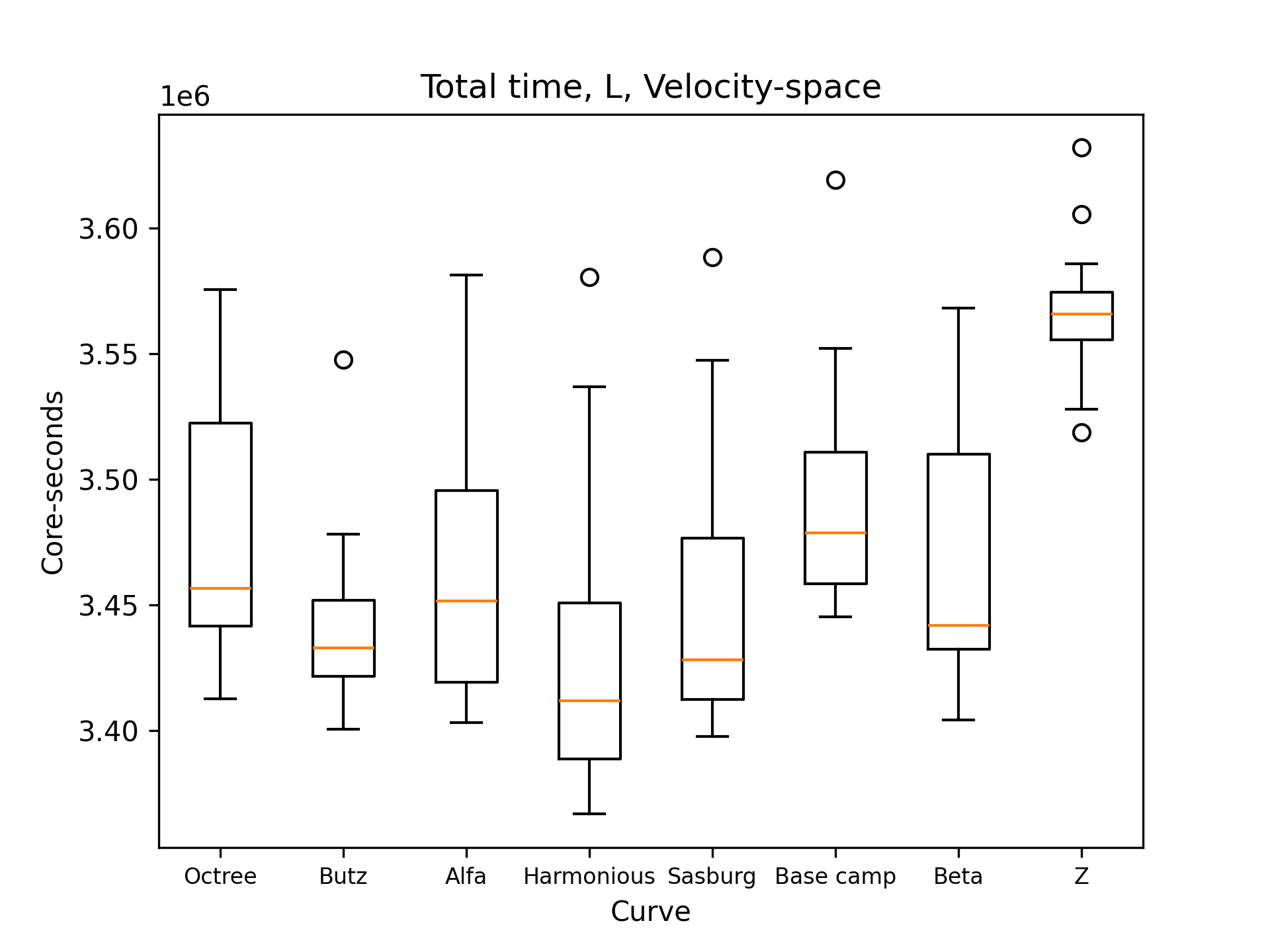}
    } \hfil
    \subfloat[]{
        \includegraphics[width=3in]{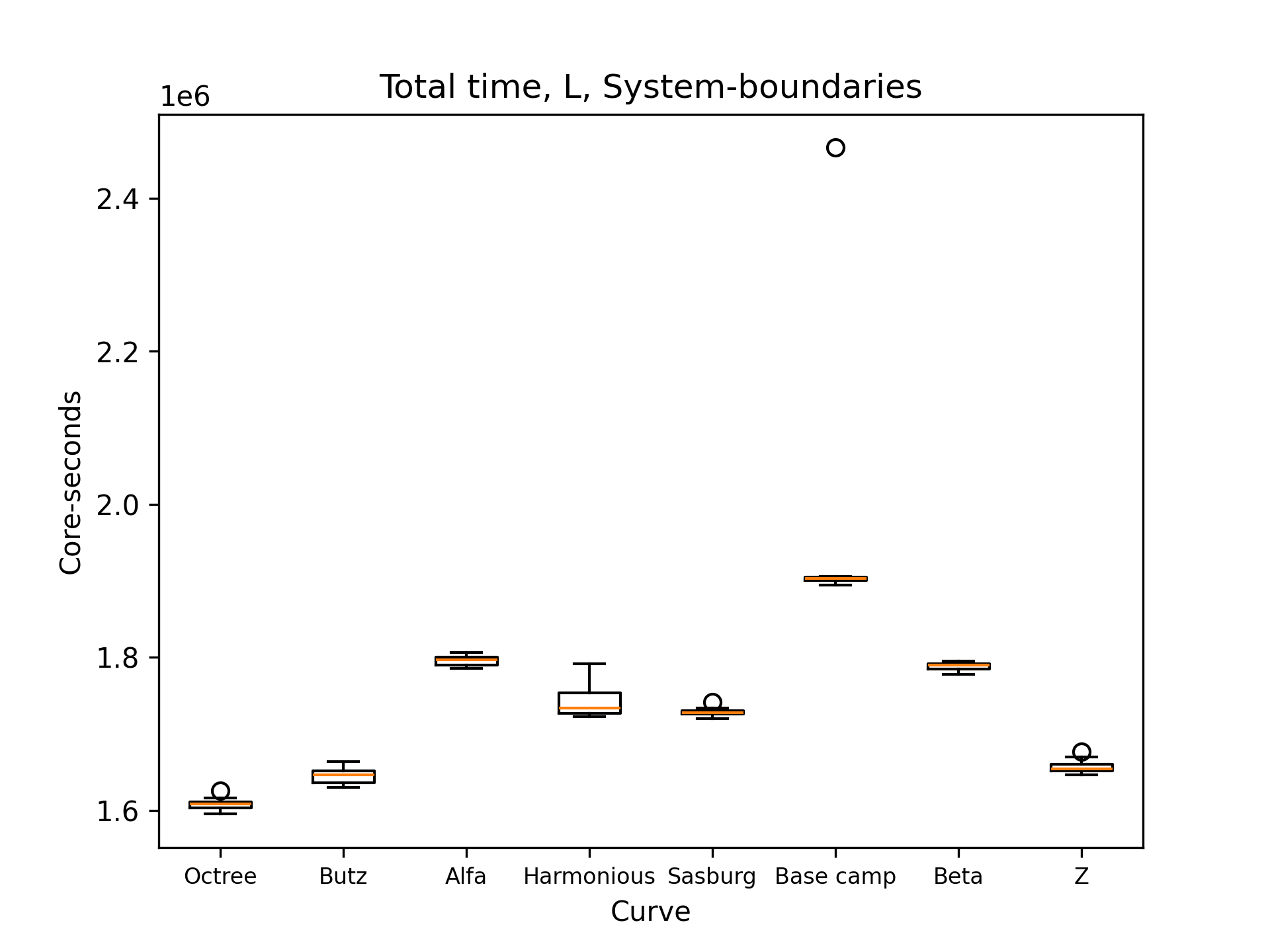}
    }
    \caption{
        Box plot of core-seconds spent in total propagation (a), spatial propagation (b), velocity space propagation (c) and system boundary updates (d) for the space filling curves tested on 12 trials of run L.
        Box plot parameters as in Figure \ref{fig:sample-timers-algs}.
        Spatial space dominates, and the Beta curve's performance there makes it the best in total time as well.
        Differences in velocity space are within error.
        This makes sense as velocity space propagation isn't related to communication between spatial cells,
        which should be the largest difference between the different curves.
    }
    \label{fig:fic-timers-hilb}
\end{figure*}

The timers for the solver sections and eight curves considered for run L are shown in Figure \ref{fig:fic-timers-hilb}.
The differences are more pronounced;
Beta had better performance than Octree by about $14\,\%$ in spatial propagation and $4\,\%$ in total time for a much better margin than for run S. 
Almost every other curve performed worse than Octree,
while one of the worst performers in the smaller test, Base Camp, had comparable performance.
As can be seen in Figure \ref{fig:mem-hilb} (b),
most curves have similar memory overhead.
Notably both the Z-curve and Base camp are significantly worse than the others.
The poor localities of these curves and particularly the discontinuity of the Z-curve explain this,
as they lead to a higher ghost cell count.

The strangest discrepancy in both tests is between the Octree and Butz curves, which should functionally identical.
This discrepancy might be explained by the two implementations having different coordinate orders.
As the magnetosphere is not symmetric and the Vlasiator simulation domain is anisotropic,
the coordinate order can have an effect on load balancing results.
Beta's consistent improvement in both tests indicates it is a better curve choice for our load balancing,
possibly due to its clearly superior $L_1$-dilation or due to an aggregate effect of all three dilations examined being better than Octree, while other curves perform better on two dilations at most.


%

\section{Conclusion} \label{sec:conclusions} 
The investigation of different algorithms revealed space-filling curves to be the optimal choice for load balancing of global Vlasiator simulations.
Since the HSFC method only explicitly optimizes for computation weight, the imbalance ends up smaller than for vertex weights in PHG partitioning,
while the excellent locality properties of Hilbert curves mean communication is optimized better than with RCB and RIB.

Further, testing different space-filling curves showed that the most common implementation of the three-dimensional Hilbert curve was not optimal.
Contrary to the initial hypothesis, this speed-up was not directly correlated to only $L_1$ dilation,
as for example the Sasburg curve with lower $L_1$-dilation performed worse than Octree.
However the Beta curve, which is superior to Octree in all locality metrics considered and has the lowest $L_1$ dilation of the six curves had a performance improvement of several percent, with the largest improvement in spatial propagation.

More generally, the two hyperorthogonal curves, Alfa and Beta, are expected to be optimal for load balancing purposes.
Alfa has optimal $L_\infty$ dilation, while Beta is superior compared to Butz in most metrics and has optimal $L_1$ and $L_2$ dilation.
These performance gains were relatively minor compared to the choice of algorithm, 
but any gains in load balance are effectively free since they don't compromise on simulation accuracy and only require testing to determine the optimal curve for the application.

For the next steps, we consider the testing of different coordinate ordering in HSFC partitioning to be a priority.
This may bridge the gap between the newly introduced curves and Zoltan's original implementation of Octree, 
as well as possibly improve further the performance lead of the Beta curve.
Another consideration is load balancing overhead and frequency.
In this work load balancing was done on static intervals and time spent in load balancing was not taken into account.
In case load balancing overhead is determined to be substantial,
measures can be taken to reduce it such as evaluating the load imbalance before deciding whether to balance.

Another consideration is hierarchical partitioning omitted from this work.
As we have established a difference in performance in parts of the simulation for the various methods,
there is a better basis of assessing whether prioritizing communication load on the inter-node level might provide a benefit.
The need for such optimizations might also become greater with the introduction of future supercomputers such as Roihu with more cores and thus processes per node \cite{roihuhw}.

\section*{Acknowledgment}
Author contributions are as follows:
\begin{enumerate}
    \item L. Kotipalo: Conceptualization; methodology; software; formal analysis; investigation; writing; visualization
    \item Markus Battarbee: Methodology; writing; supervision; project administration
    \item Yann Pfau-Kempf: Methodology; resources; writing; supervision; funding acquisition
    \item Vertti Tarvus: Investigation; resources; data curation
    \item Minna Palmroth: Supervision; project administration; funding acquisition
\end{enumerate}

The simulations and data analysis presented was done on the EuroHPC LUMI supercomputer.
The authors wish to thank the University of Helsinki local computing infrastructure and the Finnish Grid and Cloud Infrastructure (FGCI) for supporting this project with computational and data storage resources. 



\bibliographystyle{IEEEtran}
\bibliography{bibliography.bib}

\end{document}